\documentclass[conference]{IEEEtran}
\IEEEoverridecommandlockouts
\usepackage{cite}
\usepackage{amsmath,amssymb,amsfonts,pifont}
\usepackage{algorithmic}
\usepackage{graphicx}
\usepackage{textcomp}
\usepackage[table]{xcolor}
\def\BibTeX{{\rm B\kern-.05em{\sc i\kern-.025em b}\kern-.08em
    T\kern-.1667em\lower.7ex\hbox{E}\kern-.125emX}}
\definecolor{groupbg}{RGB}{200,220,240}
\definecolor{rowbg}{HTML}{F7F9FC}
\definecolor{headerbg}{HTML}{E3E6EB}
\definecolor{highlightbg}{RGB}{180,220,250} 
\usepackage[caption=false,font=footnotesize]{subfig}
\usepackage{tabularx}          
\usepackage{siunitx}
\usepackage{threeparttable}
\usepackage{makecell}
\usepackage{marvosym} 
\newcommand{\euro}{\text{\EUR{}}} 
\usepackage{hyperref} 
\usepackage{tikz}
\usetikzlibrary{shapes, shadows, positioning, fit, arrows.meta}
\usepackage{pgfplots}
\pgfplotsset{compat=1.18}
\usepackage{csquotes}
\usepackage{xurl}

\colorlet{backgroundcol}{cyan!10!white}
\newcommand{\highlight}[1]{%
	\par\noindent
	\fcolorbox{black}{backgroundcol}{%
		\parbox{\dimexpr\linewidth-2\fboxsep\relax}{%
			#1
		}%
}}

\definecolor{myblue}{RGB}{37,165,203}
\definecolor{FAUblue}{rgb}{0.000, 0.2196, 0.3961}
\definecolor{myred}{RGB}{175,32,67}
\newcommand{\AAadd}[1]{{\color{black}#1}}
\newcommand{\AAcomm}[1]{{\color{violet}#1\color{myred}}}
\renewcommand{\AAcomm}[1]{#1}

\begin{document}

\title{Wattlytics: A Web Platform for Co-Optimizing Performance, Energy, and TCO in HPC Clusters}

\author{
	\IEEEauthorblockN{Ayesha Afzal}
	\IEEEauthorblockA{\textit{Erlangen National High Performance} \\
		\textit{Computing Center (NHR@FAU)}\\
		Erlangen, Germany \\
		0000-0001-5061-0438}
	\and
	\IEEEauthorblockN{Georg Hager}
	\IEEEauthorblockA{\textit{Erlangen National High Performance} \\
		\textit{Computing Center (NHR@FAU)}\\
		Erlangen, Germany \\
		0000-0002-8723-2781}
	\and
	\IEEEauthorblockN{Gerhard Wellein}
	\IEEEauthorblockA{\textit{Department of Computer Science,} \\
		\textit{Friedrich-Alexander-Universität}\\
		Erlangen, Germany \\
		0000-0001-7371-3026}
}

\maketitle

\begin{abstract}

\AAadd{The escalating computational demands and energy footprint of GPU-accelerated computing systems complicate informed design and operational decisions. We present the first release of \textsf{Wattlytics\footnote{\label{foot:wattlytics}Web platform: \url{https://wattlytics.netlify.app}; underlying source code: \url{https://github.com/AyeshaAfzal91/PerfPerTCO}.}}, an interactive, browser-based decision-support system.
Unlike existing procurement-oriented calculators, Wattlytics integrates benchmark-driven GPU performance scaling, dynamic voltage and frequency scaling (DVFS)-aware piecewise power modeling, and multi-year total cost of ownership (TCO) analysis within a single interactive environment.
Users can configure heterogeneous systems across contemporary GPU architectures (GH200, H100, L40S, L40, A40, A100, and L4), select representative scientific workloads (e.g., GROMACS, AMBER), and explore deployment scenarios under constraints such as energy prices, system lifetime, and frequency scaling. 
Wattlytics computes multidimensional decision metrics (TCO breakdown, work-per-TCO, power-per-TCO, and work-per-watt-per-TCO) and supports design-space exploration, what-if scenarios, sensitivity metrics (elasticity, Sobol indices, Monte Carlo) and 
collaborative features to guide realistic cluster design and procurement under uncertainty. We validate the fitted power and performance models and compare four deployment modes: \emph{fixed budget}, \emph{GPU count}, \emph{performance}, and \emph{power}. 
Our case studies show that, under budget or energy constraints, optimally deployed energy-efficient GPUs can outperform higher-performance alternatives in overall cost-effectiveness. 
Wattlytics helps users explore the design parameter space and distinguish between cost- and risk-driving factors, turning HPC design into a well-informed and explainable decision-making process.
}

\end{abstract}


\section{Introduction}
High-performance computing (HPC) has entered a GPU-centric era, driven by the escalating demands of scientific simulation, machine learning, and data-intensive analytics. Successive GPU generations 
have delivered remarkable performance gains, albeit at the cost of increasing energy consumption, acquisition complexity and economic volatility. Consequently, HPC stakeholders face a multidimensional optimization challenge: designing and operating GPU systems that balance performance, energy consumption, and total cost of ownership (TCO) across multi-year deployments~\cite{green500}.

\paragraph*{Problem statement} Despite growing emphasis on energy efficiency and decarbonization, design and procurement workflows frequently rely on isolated metrics such as vendor peak specifications, standalone benchmark scores, or heuristic TCO estimates~\cite{SHAO:2022,Klemick:2019}. Fragmented approaches neglect the coupled influences of workload characteristics, frequency-dependent performance and power scaling, variable electricity pricing, and long-term operational costs, often leading to suboptimal or unsustainable decisions\AAcomm{; for molecular dynamics (MD), the most cost-efficient GPUs are not the fastest~\cite{Kutzner:2015,Kutzner:2019}}.

\paragraph*{Research gap} Existing tools typically model only one dimension of this complex landscape (performance, power, or cost). 
To our knowledge, no publicly available decision-support platform simultaneously (i) integrates benchmark-driven GPU performance models under frequency scaling~\cite{Fan:2020}, (ii) couples them with DVFS-aware power estimators~\cite{Guerreiro:2018,MEI:2017}, and (iii) embeds them into a flexible multi-year TCO model with uncertainty quantification and sensitivity analysis. Only such integration can answer \emph{how robust a GPU ranking is, and which assumption would overturn it}.

\paragraph*{Proposed solution} To address this gap, we introduce \textsf{Wattlytics}\footref{foot:wattlytics},
an interactive, browser-based decision-support platform that unifies these models. Users specify system configurations (e.g., GPU count, GPUs per node, baseline power, GPU frequencies), workload profiles (GROMACS, AMBER, or custom uploads), and economic parameters (e.g., capital cost, energy price, lifetime, operating expenses); Wattlytics evaluates TCO breakdown, work-per-TCO, power-per-TCO, and work-per-watt-per-TCO across frequencies and deployment strategies, together with sensitivity metrics. 

\paragraph*{Contributions} The key contributions of this work are: 
\begin{enumerate}
    \item \emph{A unified decision-support system} that integrates benchmark-based performance scaling, DVFS-informed power models, and configurable multi-year TCO accounting, \AAcomm{enabling reproducible trade-off exploration.}
    \item \emph{A robust platform for decision-making under uncertainty and what-if scenarios} quantifying how GPU count, frequency, energy cost, system lifetime, and deployment strategies impact work-per-TCO and efficiency metrics.
    \item \emph{Systems-level case studies} revealing non-intuitive design trade-offs across heterogeneous GPU architectures, where energy-efficient, lower-performance GPUs can be more cost-effective under realistic constraints.
\end{enumerate}

\begin{table}[t]
\caption{Capability matrix of representative tools and Wattlytics. \\
$\checkmark$ = supported, $\triangle$ = partially supported, $\times$ = not supported.}
\centering
\renewcommand{\arraystretch}{1}
\rowcolors{2}{rowbg}{white}
\begin{tabular}{|>{\centering\arraybackslash}p{4.1cm}|>{\centering\arraybackslash}p{0.4cm}|>{\centering\arraybackslash}p{0.52cm}|>{\centering\arraybackslash}p{0.4cm}|>{\centering\arraybackslash}p{1.35cm}|}
\hline
\rowcolor{headerbg}
\textbf{Tool} & \textbf{Perf.} & \textbf{Power} & \textbf{TCO} & \textbf{Scope} \\
\hline
AccelWattch~\cite{accelwattch} & $\checkmark$ & $\checkmark$ & $\times$ & GPU-only \\
PowerSensor3~\cite{powersensor3} & $\times$ & $\checkmark$ & $\times$ & Node \\
EAR~\cite{ear} & $\times$ & $\checkmark$ & $\times$ & Node \\
Accel-Sim~\cite{accelsim} & $\checkmark$ & $\times$ & $\times$ & GPU-only \\
Powerlog~\cite{powerlog} & $\times$ & $\checkmark$ & $\times$ & GPU-only \\
LIKWID~\cite{likwid} & $\triangle$ & $\triangle$ & $\times$ & CPU-only \\
TAU, HPCToolkit~\cite{hpctoolkit} & $\checkmark$ & $\times$ & $\times$ & CPU/GPU \\
AIMeter~\cite{watsonai} & $\times$ & $\checkmark$ & $\times$ & Node/Cloud \\
WattScope~\cite{wattscope} & $\times$ & $\triangle$ & $\times$ & Node \\
CodeCarbon~\cite{benoit_courty_2024} & $\times$ & $\checkmark$ & $\times$ & Node/Cloud \\
Koomey et al.~\cite{koomey2007_tco} & $\times$ & $\times$ & $\checkmark$ & System \\
TCO (NVIDIA~\cite{nvidia_tco}, Intel~\cite{intel_xeon_advisor}, AMD~\cite{amd_epyc_tco}, Scale~\cite{scale_tco}) & $\times$ & $\triangle$ & $\checkmark$ & Node/Rack \\
AWS~\cite{aws_calculator}, Google~\cite{google_calculator}, Azure~\cite{azure_calculator} & $\times$ & $\times$ & $\checkmark$ & Cloud \\
Cloud Carbon Footprint~\cite{cloudcarbonfootprint} & $\times$ & $\checkmark$ & $\times$ & Cloud/Node \\
DC Pro~\cite{dcpro} & $\times$ & $\triangle$ & $\times$ & Node/System \\
LT-TCO~\cite{lttco}, IPACK~\cite{ipack2007}, SCE/TCO~\cite{spie2007} & $\times$ & $\times$ & $\checkmark$ & Node/System \\
SPEC Power~\cite{specpower}, SERT~\cite{specsert}, Green500~\cite{green500}, MLPerf~\cite{mlperfpower} & $\triangle$ & $\checkmark$ & $\times$ & Node/System \\
LCA of HPC centres~\cite{wadenstein2025lca} & $\times$ & $\triangle$ & $\triangle$ & System \\
Energy-price TCO~\cite{aritz2025} & $\times$ & $\triangle$ & $\checkmark$ & System \\
\rowcolor{highlightbg}
Wattlytics (proposed) & $\checkmark$ & $\checkmark$ & $\checkmark$ & System \\
\hline
\end{tabular}
\label{tab:relatedwork}
\vspace{-0.2cm}
\end{table}


\paragraph*{Roadmap} The remainder of this paper is structured as follows: Section~\ref{sec:related} surveys related tools and positions Wattlytics in this landscape. Section~\ref{sec:setup} describes the experimental setup. Section~\ref{sec:design} presents the architecture, including performance, power, TCO, and sensitivity models. Section~\ref{sec:evaluation} presents case studies and quantitative insights \AAcomm{(Q1--Q9).} Section~\ref{sec:conclusion} concludes and outlines directions for future work.

\section{Related Work} \label{sec:related}
Modeling the performance, power, and cost of HPC systems spans multiple methodological domains. Existing frameworks typically emphasize a single dimension (performance modeling, power measurement, or cost analysis) while rarely integrating all three. 
\AAadd{Table~\ref{tab:relatedwork} summarizes representative tools and positions Wattlytics within this landscape}. 

\subsubsection{GPU power modeling frameworks}
\textsf{AccelWattch}~\cite{accelwattch}, built on GPGPU-Sim and Accel-Sim~\cite{accelsim}, provides cycle-level GPU power modeling with DVFS and gating effects but is impractical for system-level or cost studies. 
\textsf{PowerSensor3}~\cite{powersensor3} offers high-frequency (up to 20~kHz) hardware-based power measurement, yet lacks integration with performance or cost frameworks.
While low-level tools target microarchitectural fidelity, higher-level schedulers such as \textsf{EAR}~\cite{ear} optimize throughput under power caps but omit TCO considerations.

\subsubsection{GPU performance and benchmarking tools}
\textsf{Accel-Sim}~\cite{accelsim} simulates CUDA workloads for performance analysis but disregards power and cost. 
\textsf{Powerlog}~\cite{powerlog} records GPU power draw via \emph{nvidia-smi} for profiling and energy estimation, yet provides no predictive or workload-coupled modeling.

\subsubsection{Performance and profiling tools}
Frameworks such as \textsf{LIKWID}~\cite{likwid}, \textsf{Perf}, \textsf{TAU}, and \textsf{HPCToolkit}~\cite{hpctoolkit} support detailed CPU profiling and affinity-aware analysis~\cite{AfzalThesis:2015,Afzal:2023:2,AfzalHW:2025:1}. TAU and HPCToolkit also profile GPUs, but none of these tools models cost.

\subsubsection{Sustainability and environmental-impact tools}
\textsf{AIMeter}~\cite{watsonai}, \textsf{WattScope}~\cite{wattscope}, and \textsf{CodeCarbon}~\cite{benoit_courty_2024} estimate the energy or carbon footprint of AI and data-center workloads by combining runtime, power, and regional energy mixes. While valuable for emissions reporting, these tools lack predictive modeling, budget-constrained optimization, or generalizable TCO analysis.

\begin{table*}[t]
\caption{Specifications and efficiency metrics of NVIDIA GPUs. Bold values indicate best or key categorical values per column.}
\centering
\renewcommand{\arraystretch}{1.2}
\setlength{\tabcolsep}{8pt} 
\rowcolors{2}{rowbg}{white}
\begin{threeparttable}
\begin{tabular}{|
>{\centering\arraybackslash}p{1cm}| 
>{\centering\arraybackslash}p{1.8cm}| 
>{\centering\arraybackslash}p{2.26cm}| 
>{\centering\arraybackslash}p{0.55cm}| 
>{\centering\arraybackslash}p{0.8cm}| 
>{\centering\arraybackslash}p{0.55cm}| 
>{\centering\arraybackslash}p{0.95cm}| 
>{\centering\arraybackslash}p{1cm}| 
>{\centering\arraybackslash}p{0.85cm}| 
>{\centering\arraybackslash}p{1.75cm}| 
>{\centering\arraybackslash}p{0.63cm}|} 
\hline
\rowcolor{headerbg}
\textbf{GPUs} & \textbf{Memory clock idle / max} & \textbf{Graphics clock min / max / step} &
\textbf{SMs} & \textbf{CUDA cores} & \textbf{TDP} & \textbf{Memory type} &
\textbf{Memory capacity} & \textbf{Release year} & \textbf{Architecture} & \textbf{Process node} \\
\rowcolor{headerbg}
$\blacklozenge$ & \makebox[0pt][c]{[GHz]} & \makebox[0pt][c]{[GHz]} & & $\ddagger$ & 
\makebox[0pt][c]{[W]} & $\dagger\dagger$ & \makebox[0pt][c]{[GB]} & & $\parallel$ & \makebox[0pt][c]{[nm]} \\
\hline
L4    & 0.405 / 6.251   & 0.21 / 2.04 / 0.015 & 60  & 7,680   & 72  & GDDR6 & 24  & \textbf{2023} & Ada Lovelace & \textbf{4} \\
A40   & 0.405 / 7.251   & 0.21 / 1.74 / 0.015 & 84  & 10,752  & 300 & GDDR6 & 48  & 2020 & Ampere & 7 \\
L40   & 0.405 / \textbf{9.001}   & 0.21 / \textbf{2.49} / 0.015 & \textbf{142} & \textbf{18,176} & 300 & GDDR6 & 48  & 2022 & Ada Lovelace & \textbf{4} \\
A100  & -- / 1.215         & 0.21 / 1.41 / 0.015 & 108 & 6,912   & 400 & HBM2  & 40  & 2020 & Ampere & 7 \\
H100  & -- / 1.593         & 0.345 / 1.98 / 0.015 & 132 & 16,896  & 700 & HBM3  & 80  & 2022 & Hopper & \textbf{4} \\
GH200 & -- / 1.593         & 0.345 / 1.98 / 0.015 & 132 & 16,896  & \textbf{900} & \textbf{HBM3e} & \textbf{96} & \textbf{2023} & \textbf{Grace Hopper} & \textbf{4} \\
\hline
\end{tabular}
\setlength{\tabcolsep}{0.1pt}
\begin{tabular}{|
>{\centering\arraybackslash}p{0.9cm}|   
>{\centering\arraybackslash}p{1.1cm}|   
>{\centering\arraybackslash}p{1.3cm}|   
>{\centering\arraybackslash}p{1.4cm}|   
>{\centering\arraybackslash}p{1.3cm}|   
>{\centering\arraybackslash}p{1.35cm}|  
>{\centering\arraybackslash}p{1.22cm}|  
>{\centering\arraybackslash}p{1.32cm}|  
>{\centering\arraybackslash}p{1.1cm}|   
>{\centering\arraybackslash}p{1.33cm}|   
>{\centering\arraybackslash}p{1.15cm}|   
>{\centering\arraybackslash}p{1.35cm}|  
>{\centering\arraybackslash}p{1.55cm}|  
>{\centering\arraybackslash}p{1.82cm}|   
}
\hline
\rowcolor{groupbg}
\multicolumn{1}{|c|}{\textbf{}} &
\multicolumn{4}{c|}{\textbf{Raw specifications}} &
\multicolumn{3}{c|}{\textbf{Energy efficiency}} &
\multicolumn{2}{c|}{\textbf{Arch. efficiency}} &
\multicolumn{3}{c|}{\textbf{Cost efficiency}} &
\multicolumn{1}{c|}{\textbf{Compos. index}} \\
\hline
\rowcolor{headerbg}
\textbf{GPUs} & \textbf{$b_\text{mem}$} & \textbf{Approx. cost tier} & \textbf{Theor. peak FP32} &
\textbf{Power cap min--max} & $\dfrac{\textbf{Power cap}}{\textbf{TDP}}$ &
$\dfrac{\textbf{FP32}}{\textbf{TDP}}$ & 
$\dfrac{\textbf{$b_\text{mem}$}}{\textbf{TDP}}$ &
$\dfrac{\textbf{FP32}}{\textbf{SM}}$ &
$\dfrac{\textbf{$b_\text{mem}$}}{\textbf{FP32}}$ &
$\dfrac{\textbf{FP32}}{\textbf{Cost}}$ &
$\dfrac{\textbf{$b_\text{mem}$}}{\textbf{Cost}}$ &
$\dfrac{\textbf{FP32}}{\textbf{TDP}\times\textbf{Cost}}$ &
$\dfrac{\textbf{FP32}\times\textbf{$b_\text{mem}$}}{\textbf{TDP}\times\textbf{Cost}}$ \\
\rowcolor{headerbg}
 & [GB/s]$\parallel$ & $\S$ & [TF]$\blacktriangle$ & [W] & [\%TDP]$\bullet$ &
(norm.)$\dagger$ & (norm.)$\Box$ & (norm.)$\lozenge$ & (norm.)$\diamond$ & (norm.)$\triangleleft$ & (norm.)$\circ$ & (norm.)$\star$ & (norm.)$\P$ \\
\hline
L4    & 300  & Low    & 30.3
~\cite{webI} & 40--72 & \textbf{56--100} & \textbf{1.00} & 0.15 & 0.79 & 0.12 & 0.72 & 0.87 & \textbf{1.00} & \textbf{1.00} \\
A40   & 798  & Medium & 37.4
~\cite{webII} & 100--300 & 33--100 & 0.29 & 0.41 & 0.69 & 0.26 & 0.42 & 0.53 & 0.40 & 0.43 \\
L40   & 864  & Medium & \textbf{90.5}
~\cite{webIII} & 100--300 & 33--100 & 0.74 & 0.45 & \textbf{1.00} & 0.12 & \textbf{1.00} & 0.60 & 0.73 & 0.74 \\
A100  & 1,555 & Medium & 19.5
~\cite{webIV} & 100--400 & 25--100 & 0.12 & 0.81 & 0.28 & \textbf{1.00} & 0.22 & 0.81 & 0.17 & 0.25 \\
H100  & 3,352 & \textbf{V. high} & 67.0
~\cite{webV}& 350--700 & 50--100 & 0.24 & 0.70 & 0.78 & 0.50 & 0.18 & \textbf{1.00} & 0.27 & 0.85 \\
GH200 & \textbf{4,000} & \textbf{V. high} & 67.0
~\cite{webVI} & \textbf{400--900} & 44--100 & 0.18 & \textbf{1.00}  & 0.78 & 0.75 & 0.18 & 0.93 & 0.23 & 0.68 \\
\hline
\end{tabular}

\begin{tablenotes}[flushleft]\footnotesize
\item[$\ddagger$] The number of CUDA cores is calculated as Streaming Multiprocessors (SMs) × cores per SM, which varies across architectures.
\item[$\dagger\dagger$] Memory type affects achievable bandwidth: GDDR6 uses wide I/O; HBM2/HBM3 are 3D-stacked; HBM3e adds improved signaling for highest bandwidth.
\item[$\parallel$] Newer Ada/Hopper/GH GPUs achieve $\sim$2–5× memory bandwidth and $\sim$3× FP32 throughput over Ampere due to denser process nodes and HBM memory. 
\item[$\S$] Qualitative retail cost tiers: low ($\leq5\,\text{k\$})\approx3.5\,\text{k\$}$, medium ($5$--$10\,\text{k\$})\approx7.5\,\text{k\$}$, high ($10$--$25\,\text{k\$})\approx17.5\,\text{k\$}$, and very high ($\geq25\,\text{k\$})\approx30\,\text{k\$}$.
\item[$\blacktriangle$] Approx. \emph{peak theoretical single-precision FP32 throughput} ($\frac{\text{CUDA Cores} \times 2 \times \text{Boost Clock [GHz]}}{1000}$) values are taken from public or official NVIDIA datasheets \cite{nvidia_specs}. 
\item[$\bullet$] Percentage of TDP equals $(\text{power cap} / \text{TDP}) \times 100$ and indicates allowable power reduction below nominal TDP.
\item All derived metrics are normalized to column maxima:
    [$\dagger$] Compute-power efficiency -- normalized FP32 performance per watt (L4 best-case due to power limiting). 
    [$\Box$] Memory energy efficiency — memory bandwidth $b_\text{mem}$ per watt.
    [$\lozenge$] Compute density per SM --  each SM contribution to total FP32 throughput. 
    [$\diamond$] Memory-compute balance -- memory bandwidth available per unit of compute performance. High A100/GH200 ratios reflect wide HBM buses.
    [$\triangleleft$] Compute-cost efficiency -- normalized FP32 throughput per cost tier.
    [$\circ$] Bandwidth–cost index -- memory bandwidth per cost tier.
    [$\star$] Composite efficiency -- combined compute, energy, and cost efficiency.
    [$\P$] GPU composite Index --  integrated metric of compute, bandwidth, power, and cost efficiency for cross-GPU comparison.

\end{tablenotes}
\end{threeparttable}
\label{tab:gpu_specs_perf}
\end{table*}

\subsubsection{TCO modeling tools}
Foundational work by \textsf{Koomey et al.}~\cite{koomey2007_tco} and the Uptime Institute decomposed costs into capital and operational components, emphasizing energy, infrastructure, and facility contributions. It provides open methods for estimating power, cooling, and IT costs, but omits performance and dynamic power effects.  
Vendor calculators (e.g., \textsf{NVIDIA TCO Calculator}~\cite{nvidia_tco}, \textsf{Intel Xeon Advisor}~\cite{intel_xeon_advisor}, \textsf{AMD EPYC Estimator}~\cite{amd_epyc_tco}, and \textsf{Scale Computing’s estimator}~\cite{scale_tco}) offer quick node- or rack-level assessments but rely on proprietary assumptions, limiting research transparency.
Academic and open-source tools, such as \textsf{Cloud Carbon Footprint}~\cite{cloudcarbonfootprint} and \textsf{DC Pro}~\cite{dcpro}, estimate energy or emissions but ignore throughput-dependent behavior at the hardware level. 
Cloud vendor pricing calculators from \textsf{AWS}~\cite{aws_calculator}, \textsf{Google Cloud}~\cite{google_calculator}, and \textsf{Microsoft Azure}~\cite{azure_calculator} support deployment cost comparisons but omit energy, cooling, or benchmark-dependent modeling.
Academic data-center cost frameworks, including \textsf{LT-TCO}~\cite{lttco}, infrastructure models~\cite{ipack2007}, and SCE/TCO algorithms~\cite{spie2007}, address long-term expenditures but generally ignore benchmark-driven performance or heterogeneous GPU scenarios.
Efficiency standards such as \textsf{SPEC Power}~\cite{specpower}, \textsf{SPEC SERT}~\cite{specsert}, \textsf{Green500}~\cite{green500}, \textsf{ENERGY STAR}~\cite{energystar}, and \textsf{MLPerf Power}~\cite{mlperfpower} measure energy-performance trade-offs across workloads and systems but omit holistic TCO models or cost-driven optimization.  
Recent studies extend TCO analysis to embodied versus operational emissions in HPC centres~\cite{wadenstein2025lca} and to the impact of electricity price volatility on long-term HPC operational costs~\cite{aritz2025}, yet integration with benchmark-based modeling remains limited.

\subsubsection{Positioning Wattlytics}
Many HPC centers rely on non-public, center-specific TCO calculators, while profilers and simulators target microarchitectural fidelity rather than system-level trade-offs.
In contrast, Wattlytics fills a critical gap between microarchitectural simulation tools and high-level procurement calculators by combining analytical gray-box modeling with accessibility.
\textsf{Wattlytics} is an open, interactive, browser-based platform that unifies three rarely combined dimensions: (i) benchmark-driven GPU performance under frequency scaling, (ii) DVFS-aware power estimation, and (iii) multi-year TCO modeling with scenario and sensitivity analysis.
This integration allows mapping workloads to heterogeneous GPU configurations under variable energy or budget constraints and provides transparent and holistic metrics such as \emph{work-per-watt-per-TCO}, bridging gaps left by generalized infrastructure models, cloud-carbon estimators, and vendor TCO tools. 

\section{Experimental setup}\label{sec:setup}

\AAadd{Wattlytics is applicable to any application class; for evaluation, we selected six GROMACS~\cite{AfzalHW:2025:2} and eleven AMBER \cite{webLI} 
benchmarks, which dominate the GPU workload mix at the HPC center where the measurements were taken, owing to their focus on atomistic molecular dynamics (MD) simulations.} The input parameter sets span small to large atom counts, with larger benchmarks exhibiting greater memory-bandwidth sensitivity. Simulations used GROMACS~2024.4 (GCC~11.2, Intel MKL, CUDA~12.4) and AMBER~2024.2 (GCC~11.2, MKL, CUDA~12.4), 
with GPU acceleration enabled for all major kernels. CPU/GPU affinity was carefully managed to ensure reproducible placement and minimize contention. \AAcomm{All profiles are single-GPU:} GROMACS runs used one MPI rank 
with 16 OpenMP threads, neighbor list updates every 20 steps, single-precision floating point, and 200,000 MD steps via \texttt{gmx mdrun} (capped at 0.2~h). The AMBER benchmarks used \texttt{pmemd.cuda -O}. 
Each configuration was repeated three times; run-to-run performance variability remained below 5\%, except for selected power-capped runs with frequent GPU frequency transitions; locked frequencies avoid such transitions. \AAadd{Performance is reported as the average simulation throughput in nanoseconds per day (ns/day) over the last 800 steps, capturing steady-state performance after GPU warm-up and avoiding underestimation.}

\begin{figure*}[t]
\centering
\resizebox{2.05\columnwidth}{!}{
\begin{tikzpicture}[font=\sffamily, >= {Latex[length=5mm, width=4mm]}, thick,
    node distance=0.7cm, align=center,
    module/.style={rectangle, rounded corners, draw=black!70, fill=blue!5, very thick, minimum width=2.8cm, minimum height=1cm},
    smallmod/.style={rectangle, rounded corners, draw=black!60, fill=blue!10, minimum width=2.6cm, minimum height=0.8cm},
    layer/.style={draw=black!40, dashed, rounded corners, fill=green!10, inner sep=0.5cm},
    arrow/.style={
        ->,
        line width=2.2pt,
        draw=black!90,
        shorten >=4pt,
        shorten <=4pt
    },
    io/.style={ellipse, draw=black!70, fill=orange!10, minimum width=2cm, minimum height=0.8cm}
]
\Huge
\node[layer, fill=orange!7, inner sep=6pt, label=below:{\bfseries\fontsize{26pt}{30pt}\selectfont Input layer}] (IP) {%
    \begin{tikzpicture}[node distance=0.25cm, align=center]
        \node[layer, fill=orange!7] (inputs)  {%
            \begin{tikzpicture}[node distance=0.25cm, align=center]
                \node[smallmod, fill=orange!14] (specs) {Hardware specs\\(type, frequency, power cap)};
                \node[smallmod, fill=orange!14, below=0.5cm of specs] (bench) {Benchmark database\\(type, ID, baseline power)};
            \end{tikzpicture}
        };

        \node[layer, fill=orange!7, below=0.4cm of inputs] (costIP)  {%
            \begin{tikzpicture}[node distance=0.25cm, align=center]
                \node[smallmod, fill=orange!14] (GPUprice) {GPU price (static or \\ live from DELTA Computer \cite{webLII})};
                \node[smallmod, fill=orange!14, below=0.5cm of GPUprice] (costs) {Capital \& operational costs};
            \end{tikzpicture}
        };
        \node[below=0.05cm of costIP.south] {Sliders with tooltips and strategy tips,\\ dropdowns or CSV/JSON import};

    \end{tikzpicture}
};

\node[layer, right=3cm of inputs, fill=cyan!5, inner sep=11pt, label=below:{\bfseries\fontsize{26pt}{30pt}\selectfont Modeling engine}] (model)  {%
    \begin{tikzpicture}[node distance=0.25cm, align=center]
        \node[smallmod] (tco) {TCO model};
        \node[smallmod, below=0.5cm of tco] (power) {Power model};
        \node[smallmod, below=0.5cm of power] (perf) {Performance model};
        \node[smallmod, below=0.5cm of perf] (sensitivity) {Sensitivity/uncertainty};
    \end{tikzpicture}
};


\node[layer, right=2cm of model, fill=teal!8, inner sep=6pt, label=below:{\bfseries\fontsize{26pt}{30pt}\selectfont Analysis engine}] (uncert)  {%
    \begin{tikzpicture}[node distance=0.25cm, align=center]
        \node[layer, right=2cm of model,fill=teal!8,] (analysis) {%
    \begin{tikzpicture}[node distance=0.25cm, align=center]
        \node[smallmod, fill=teal!18] (what-if) {What-if scenario analysis};
        \node[smallmod, fill=teal!18, below=0.5cm of what-if] (TCObreak) {Frequency \& power cap tuning};
        \node[smallmod, fill=teal!18, below=0.5cm of TCObreak] (metrics) {Decision metrics: TCO breakdowns, \\ work-per-TCO, power-per-TCO, \\  work-per-watt-per-TCO};
        \node[smallmod, fill=teal!18, below=0.5cm of metrics] (approaches) {Deployment strategies:\\ fixed-budget, fixed-performance,\\  fixed-power, or fixed-GPU count};
    \end{tikzpicture}
    };
    \node[below=0.05cm of analysis.south] {Per GPU and cross-GPU analyses};
    \end{tikzpicture}
};

\node[layer, right=2cm of uncert, inner sep=6pt, label=below:{\bfseries\fontsize{26pt}{30pt}\selectfont Visualization layer}] (OP) {%
    \begin{tikzpicture}[node distance=0.25cm, align=center]
        \node[layer, right=2cm of uncert] (outputs) {%
    \begin{tikzpicture}[node distance=0.25cm, align=center]
        \node[smallmod, fill=green!20] (plots) {Plots: heatmaps, bar,\\ pie, stacked, tornado\\(high-resolution PNG export)};
        \node[smallmod, fill=green!20, below=0.5cm of plots] (tables) {Sortable tables \\ (CSV export)};
        \node[smallmod, fill=green!20, below=0.5cm of tables] (reports) {Summary \& comparison reports\\ (PDF export)};
    \end{tikzpicture}
    };
    \node[below=0.05cm of outputs.south] {Dashboard toggle between \\a light mode and a dark mode};
    \end{tikzpicture}
};


\node[layer, fill=red!5, right=2cm of OP, minimum height=9cm, inner sep=6pt, label=below:{\bfseries\fontsize{26pt}{30pt}\selectfont Collaboration layer}] (collaboration) {%
    \begin{tikzpicture}[node distance=0.25cm, align=center]
        \node[smallmod, fill=red!10] (metrics) {Sharing feature: \\ embedded instant or\\ persistent share links};
        \node[smallmod, fill=red!10, below=0.5cm of metrics] (plots) {Auto-generated \\ blog summaries};
    \end{tikzpicture}
};
\node[above=0cm of collaboration.south] {FAIR principles};

\draw[arrow] (IP) -- (model);
\draw[arrow] (model) -- (uncert);
\draw[arrow] (uncert) -- (OP);
\draw[arrow] (OP) -- (collaboration);

\draw[arrow, dashed]
    (OP.north) -- ++(0,3cm)                     
    -- ++(-48cm,0) node[midway, below] {\fontsize{26pt}{30pt}\selectfont\textbf{Real-time user-driven feedback}}  
    -- ++(0,-2cm);                                   

\end{tikzpicture}
}
\caption{Wattlytics architecture showing the end-to-end pipeline from input to decision analytics. Users provide hardware, benchmark, and cost data, which feed into the Modeling Engine. The Analysis Engine quantifies uncertainties, performs frequency tuning, evaluates deployment strategies, and generates key metrics. Outputs are visualized in interactive dashboards and shared via FAIR-aligned collaboration features.}
\label{fig:dataflow}
\end{figure*}
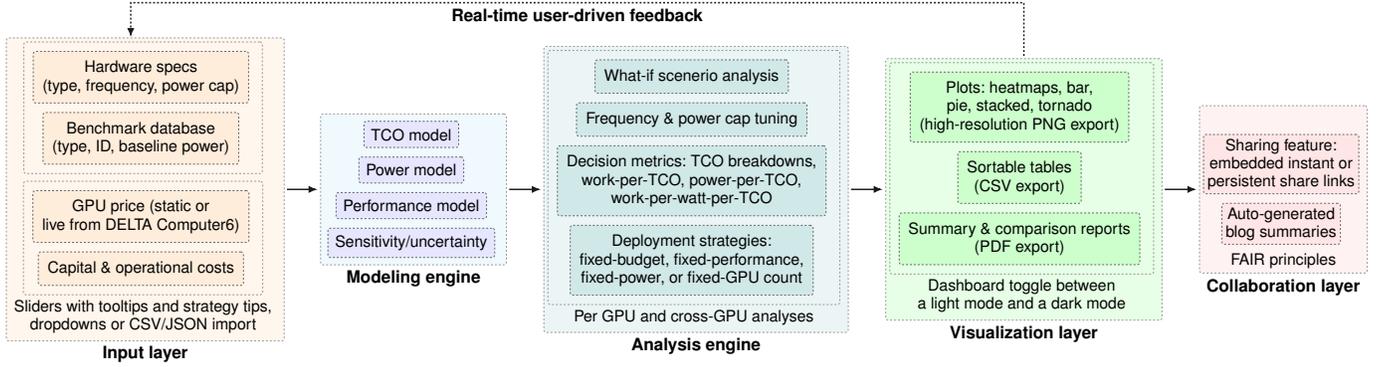

Experiments employed NVIDIA GPUs spanning Ampere to Ada Lovelace, Hopper, and Grace Hopper architectures and an AMD MI300X on production HPC clusters\footref{sec:URLanonymity1} \footref{sec:URLanonymity2} and test cluster\footnote{\url{https://doc.nhr.fau.de/clusters/testcluster}}. Table~\ref{tab:gpu_specs_perf} summarizes architectural features (such as clock ranges, SMs, CUDA cores, \AAadd{Thermal Design Power (TDP)}, memory hierarchy) and derived metrics (such as FP32 throughput per watt, per cost, per TDP). These metrics support Wattlytics' multi-objective \emph{power-per-TCO} and \emph{work-per-TCO} models. \AAcomm{A100/H100 are SXM modules, GH200 uses NVLink-C2C, and the others are PCIe cards; TDP is the default board power limit.} GPU memory clocks were fixed. Frequency sweeps \AAadd{were performed over predefined graphics clocks using \texttt{nvidia-smi} via \texttt{--applications-clocks} (older GPUs) or \texttt{--lock-gpu-clocks} (H100 and newer)\AAcomm{, or \texttt{rocm-smi} (AMD)}, conducted under the default maximum power limit (i.e., without artificial power capping)}. Binaries were compiled using architecture-specific \texttt{nvcc} optimization flags, and execution times were measured using CUDA events. Power-capping experiments employed \texttt{--power-limit} at default GPU clocks. \AAadd{Power was read from the board sensor via \texttt{nvidia-smi} every 100\,ms; samples within a steady-state window (after thermal and power stabilization, GPU utilization $>$ 80\%) are arithmetically averaged to suppress short-term DVFS fluctuations, and energy is average power times runtime.} \AAcomm{For GH200, only the Hopper GPU is metered; Grace CPU and memory count as node baseline power.} Power-capping ranges correspond to software-configurable power limits; idle memory frequency (405~MHz) was excluded, as it produces unrealistically low bandwidth for workloads. All outputs were stored to ensure reproducibility.  

\section{Wattlytics Design and Architecture}\label{sec:design} 
Wattlytics employs a \emph{modular, two-tier architecture} \AAcomm{running client-side in the browser (JavaScript)} and emphasizing \emph{user accessibility} and \emph{analytical modeling}: (i) an \emph{interactive web front-end} integrating input, analysis, visualization, and collaboration layers for rapid exploration, and (ii) a \emph{modeling back-end} with benchmark-driven analytical models\AAcomm{; together, these form the five layers of Figure~\ref{fig:dataflow}}. 

\subsection{User Interface} 
The web-based UI allows users to configure hardware, workloads, and various costs while supporting analysis, visualization, and FAIR-aligned (Findable, Accessible, Interoperable, Reproducible) collaboration without any local software installation. 

\subsubsection{Input layer}  
Wattlytics accepts a comprehensive set of user-defined inputs via sliders, dropdowns, or CSV/JSON import.
Users specify deployment strategies limiting \emph{budget}, \emph{GPU count}, \emph{performance}, and \emph{power}. \emph{Capital costs} cover nodes, servers, infrastructure, facilities, and software, providing a detailed breakdown of cost distribution across system components.
\emph{Operational costs} include electricity (\euro /kWh), Power Usage Effectiveness (PUE), node maintenance (\euro /year), system usage (hrs/year), depreciation (\euro /year), software subscription (\euro /year), and utilization inefficiency (\euro /year), enabling realistic estimation of annual operating expenses. 
These costs also include a sustainability subset covering renewable energy, decarbonization, and heat reuse revenue and factor. 
\AAadd{CO$_2$ emissions are currently excluded to avoid over-engineering, as some centers (including ours) use green electricity. Wattlytics will later include Total Carbon of Ownership per component (GPU, CPU, memory), factoring in embodied carbon and grid intensity.
For GPUs with similar work-per-TCO, carbon footprint can guide selection, favoring slightly higher-cost options with lower emissions.} Uncertainty sliders allow exploration of variable assumptions (Section~\ref{sec:uncertainty}). Power/performance-model-dependent parameters, such as node baseline power without GPUs (capturing system-level power and cost overheads), GPU frequency/power caps and node/GPU efficiency, are grouped separately.
Hardware parameters include GPU type, cost, and devices per node.
Users can specify \emph{input uncertainty for Sobol or Monte Carlo analysis} either globally for all parameters or individually for each parameter.
Each parameter $x_i$ is sampled from a uniform distribution over its uncertainty range:
\begin{equation}
x_i^{(n)} = x_i (1 + \varepsilon_i^{(n)}), \qquad \varepsilon_i^{(n)} \sim \mathcal{U}(-u_i, u_i).
\end{equation}
\emph{Benchmark parameters} define preloaded workload profiles (e.g., GROMACS, AMBER) along with their performance and power draw. Custom uploads with different applications \AAcomm{and hardware} are possible \AAcomm{(Section~\ref{sec:extensibility})}.
Tooltips, strategy tips and cluster presets (from Top500/Green500 lists) guide users; currently available profiles include
ClusterA\footnote{\label{sec:URLanonymity1}\url{https://doc.nhr.fau.de/clusters/fritz}} (A40, A100) and ClusterB\footnote{\label{sec:URLanonymity2}\url{https://doc.nhr.fau.de/clusters/helma}} (H100).
A GPU price history plot compares live market prices $C_{\text{live}}(t)$ (e.g., from the DELTA Computer website \cite{webLII}) with static baseline prices $C_{\text{static}}$, illustrating historical pricing trends and market-driven shifts over time.
The relative GPU cost difference $\Delta C_{\%}(t)$ at time $t$ is computed as follows: 
\begin{equation}
\Delta C_{\%}(t) = \frac{C_{\text{live}}(t) - C_{\text{static}}}{C_{\text{static}}} \times 100\,.  
\end{equation}

\subsubsection{Analysis engine}
Given hardware, benchmark, and cost parameters, the analysis engine computes the maximum number of GPUs purchasable within a budget, per-GPU and system TCO breakdowns, aggregate performance and energy over the system lifetime, optimal deployment strategies across GPU types 
and workloads, and parameter sensitivity. 
Outputs M are expressed as functions of the $n$ input parameters (defined in the input layer):
\begin{equation}
\mathrm{M} = f(\mathbf{x}), \quad \mathbf{x} = [x_1, x_2, \dots, x_n]
\end{equation}
where $x_i$ is the $i$-th input parameter and M can represent TCO, work-per-TCO, power-per-TCO, or work-per-watt-per-TCO.
\begin{equation}
    \text{work-per-TCO} = \frac{Q_{\text{total}}}{\text{TCO}}, \quad
    \text{power-per-TCO} = \frac{W_{\text{total}}}{\text{TCO}}, 
\end{equation}
\begin{equation}
    \text{work-per-watt-per-TCO} = \frac{Q_{\text{total}}}{W_{\text{total}} \cdot \text{TCO}}.
\end{equation}
\enquote{work-per-TCO} quantifies the computational work $Q_{\text{total}}$ delivered per unit cost, with higher values indicating superior cost effectiveness.
\enquote{Power-per-TCO} captures \AAcomm{sustained average} power $W_{\text{total}}$ per unit cost, where lower values denote more power-efficient deployments.
\enquote{work-per-watt-per-TCO} integrates performance, energy, and cost, expressing the amount of work delivered per watt and per \euro, enabling multi-dimensional evaluation of HPC system efficiency; it is analogous to the ``energy-delay product'' in energy efficiency analysis of compute devices.
\AAcomm{Since TCO already prices energy and cooling, both power-based metrics are secondary views for binding power limits, separating configurations of similar TCO.}
Side-by-side what-if comparisons highlight input and output differences between two configurations. 
Four deployment strategies are supported:
\paragraph{\emph{Fixed budget} ($B \le \text{cap}$)} Common in academia; maximizes performance within a lifetime TCO envelope.
\paragraph{\emph{Fixed GPU count} ($n_\text{GPU} \le \text{cap}$)} Same count for every GPU type, uneven spending; for limits from rack space, allocation policies, supply, or quotas.
\paragraph{\emph{Fixed performance} ($P_\text{total} \ge \text{cap}$)} Meets a throughput target at minimum cost or power, for production planning.
\paragraph{\emph{Fixed power} ($W_\text{total} \le \text{cap}$)} Performance and cost within a power or thermal envelope, for energy-limited sites and power-supply planning. \AAcomm{Each constraint can be combined with the budget cap (Section~\ref{sec:evaluation}(Q4)).}

\subsubsection{Visualization layer} 
In Wattlytics, parameters can be adjusted with immediate visual feedback, enabling iterative exploration.
Results update interactively in bar, stacked, and pie charts, sortable tables, performance/power heatmaps, and performance- and power-frequency plots, with zoom, pan, and box/lasso selection; charts and tables export as PNG/CSV, and summary and comparison reports as PDF. Sensitivity results appear as per-GPU bar plots and cross-GPU heatmaps (parameters $\times$ GPUs).
Elasticity values are GPU-specific and signed (blue for reductions, red for increases), whereas Sobol and Monte Carlo values are normalized (0--100\%) \emph{per parameter across all GPUs}, emphasizing relative GPU-specific contributions rather than absolute magnitudes.

\subsubsection{Collaboration layer}
Every configuration and result in Wattlytics can be exported, cited, and shared via \emph{instant} client-side LZ-String \cite{webLIII} compressed URLs (used in the Sitography~\cite{web1}--\cite{web11}) or, for URLs ${\gtrsim}2000$ characters, \emph{persistent} serverless Supabase links. 
To further streamline dissemination, Wattlytics auto-generates Markdown summaries embedding these links, allowing results to be instantly replicated, compared, and published across collaborative or public platforms. 

\subsubsection{Extensibility}\label{sec:extensibility}
\AAcomm{Adding a GPU of any vendor or a workload needs no code changes: users upload a JSON/CSV file (any exported configuration serves as a template) with hardware constants (price, TDP, GPUs per node, clock bounds) and throughput and power versus clock, measured via \texttt{nvidia-smi}, \texttt{rocm-smi}, or external meters (e.g., on cloud or early-access systems), to which Eq.~\eqref{eq:piecewise_power} is fitted. Each workload with distinct frequency scaling needs its own profile (Section~\ref{sec:accuracy}).}

\subsection{Modeling engine}
Wattlytics integrates four model families into a unified analytical engine\AAcomm{: fitted power and performance models (validated in Section~\ref{sec:accuracy}) and analytical TCO and sensitivity models that propagate inputs and their uncertainty}.
The models couple through $f_\text{GPU}$ and $n_\text{GPU}$: $f_\text{GPU}$ sets per-GPU power and throughput, power enters TCO via energy cost, the deployment constraint then fixes $n_\text{GPU}$, and the decision metrics combine all three.

\subsubsection{Total Cost of Ownership (TCO) modeling}
The \emph{TCO} quantifies the cumulative cost of acquiring and operating a GPU-accelerated HPC system over its lifetime:
\begin{equation}
\text{TCO} = C_\text{cap} + C_\text{op},
\end{equation}
where $C_\text{cap}$ is the one-time capital expenditure and $C_\text{op}$ is the cumulative operational expenditure over the cluster lifetime $T_\text{life}$.
For a cluster with $n_\text{GPU}$ GPUs and $g_\text{node}$ GPUs per node, the capital cost is
\begin{equation}
C_{\text{cap}} = n_{\text{GPU}} \biggl(C_{\text{GPU}} + \frac{C_{\text{ns}} + C_{\text{ni}} + C_{\text{nf}}}{g_\text{node}}\biggr) + C_{\text{sw}},
\end{equation}
where $C_{\text{GPU}}$, $C_{\text{ns}}$, $C_{\text{ni}}$, $C_{\text{nf}}$, and $C_{\text{sw}}$ denote GPU, server, infrastructure (e.g., cooling and power delivery), facility, and software costs, respectively.
Users can switch between static GPU pricing, reflecting \AAadd{the latest hardware quotes provided to our local computing center}, and live-delta pricing, which updates dynamically based on current market data. The annual operational cost $C_{\text{op,yr}}$ over the system lifetime $T_\text{life}$ is
\begin{equation}
C_{\text{op,yr}} = n_{\text{GPU}} \cdot C_\text{var} + C_\text{base}, \quad
C_{\text{op}} = T_\text{life} \cdot C_{\text{op,yr}},
\end{equation}
with variable and baseline annual costs defined as
\begin{equation}
\begin{aligned}
C_\text{var} &= \text{PUE} \cdot \Bigl(C_\text{elec} - f_\text{hr} C_\text{hr}\Bigr) \cdot 
\Bigl(\frac{W_\text{base} U_\text{sys}}{g_\text{node}} + W_\text{GPU} U_\text{sys}\Bigr)\\
    & \quad + \frac{C_\text{mnt}}{g_\text{node}},\\
 C_\text{base}  &= C_\text{dep} + C_\text{sub} + C_\text{ineff}.
\end{aligned}
\end{equation}
Here, $W_\text{base}$ and $W_\text{GPU}$ denote node baseline power (excluding GPUs) and GPU power, respectively; $U_\text{sys}$ is system utilization; PUE is the Power Usage Effectiveness; $C_\text{elec}$ is the electricity cost; $f_\text{hr}$ is the fraction of recoverable heat; $C_\text{hr}$ is the corresponding heat recovery value; and $C_\text{mnt}$ is the maintenance cost.
The cost components $C_{\text{dep}}$, $C_{\text{sub}}$, and $C_{\text{ineff}}$ denote depreciation, software subscription, and utilization inefficiency; their sum $C_\text{base}$ is invariant with system scale (e.g., facility overhead, networking, administration) and determines whether TCO scales proportionally with system size, which matters when comparing small and large deployments.
By explicitly modeling both fixed and variable annual cost components, Wattlytics provides a more realistic representation of long-term expenditure.

\subsubsection{Power modeling} 
Wattlytics derives average GPU and system power from TDP and measured frequency scaling behavior~\cite{dvfsmodel}. 
The total energy consumption over the system lifetime is computed as:
\begin{equation}
\begin{aligned}
    E_{\text{total}} &= W_{\text{system}} \cdot U_\text{sys} \cdot T_\text{life} \cdot \text{PUE}, \\
    W_{\text{system}} &= n_\text{nodes} W_{\text{base}} + n_{\text{GPU}} \cdot W_{\text{GPU}}(f_{\text{GPU}}), \\
    W_{\text{GPU}}(f_{\text{GPU}}) &= \min \big(W_{\text{TDP}}, \phi(f_{\text{GPU}})\big),
\end{aligned}
\end{equation}
where $W_{\text{GPU}}$ is the per-GPU average power, $f_\text{GPU}$ denotes the GPU core (\emph{graphics}) clock, and $W_{\text{base}}$ represents node baseline power (CPU and memory; cooling enters via PUE).
The vendor-specified thermal design power $W_{\text{TDP}}$ serves as a practical upper bound for sustained GPU power and aligns with the maximum enforceable device \emph{power limit}.
The dynamic GPU power consumption is modeled relative to a reference frequency using a piecewise linear–quadratic model, capped by $W_{\text{TDP}}$. 
The DVFS behavior~\cite{Amati:2025} is captured by a piecewise function $\phi(f_{\text{GPU}})$:
\begin{equation}
\phi(f_{\text{GPU}}) =
\begin{cases}
    b_1 f_{\text{GPU}} + c_1, & f_{\text{GPU}} \le f_t, \\
    a_2 f_{\text{GPU}}^2 + b_2 f_{\text{GPU}} + c_2, & f_{\text{GPU}} > f_t,
\end{cases}
\label{eq:piecewise_power}
\end{equation}
where $f_t$ denotes the transition between the low-frequency, leakage-dominated (linear) regime and the high-frequency, voltage-dominated (quadratic) regime.
Coefficients $\boldsymbol{\theta} = [a_2, b_1, b_2, c_1, c_2, f_t]$ are fitted via nonlinear least-squares using the Levenberg–Marquardt algorithm: 
\begin{equation}
    \min_{\boldsymbol{\theta}} \sum_{i=1}^{N} \big(W_{\text{GPU}}^{\text{measured}}(f_i) - W^{\text{model}}_{\text{GPU}}(f_i; \boldsymbol{\theta})\big)^2.
\end{equation}
The algorithm combines the stability of gradient descent and the rapid convergence of the Gauss–Newton method, providing robust and accurate fits even for nonlinearly parameterized models. The breakpoint $f_t$ is automatically determined by the fit, with an initial guess at the midpoint of the measured frequency range. For numerical smoothness, continuity at $f_t$ is encouraged via the constraint $b_1 f_t + c_1 = a_2 f_t^2 + b_2 f_t + c_2$.
\AAadd{Fit errors and their effect on decision metrics are quantified in Section~\ref{sec:accuracy}.}
Wattlytics models baseline (idle) power by extrapolating GPU power to zero frequency within the low-frequency linear regime.
This yields baseline power levels of approximately 15--20\% of TDP for the evaluated devices. In contrast, CPUs typically exhibit substantially higher baseline (idle) power than GPUs~\cite{Afzal:2023:2}.
In Wattlytics, default coefficients $\boldsymbol{\theta}$ are preloaded for standard workloads (viewable via the ``Show Power Model'' button), but users can adjust them to model alternative or hypothetical workloads. 

\begin{figure}[t]
    \centering   
    \subfloat[GROMACS Performance\label{fig:fl_md1_berendsen}]{
         \hspace{-0.1em}\includegraphics[width=0.24\textwidth]{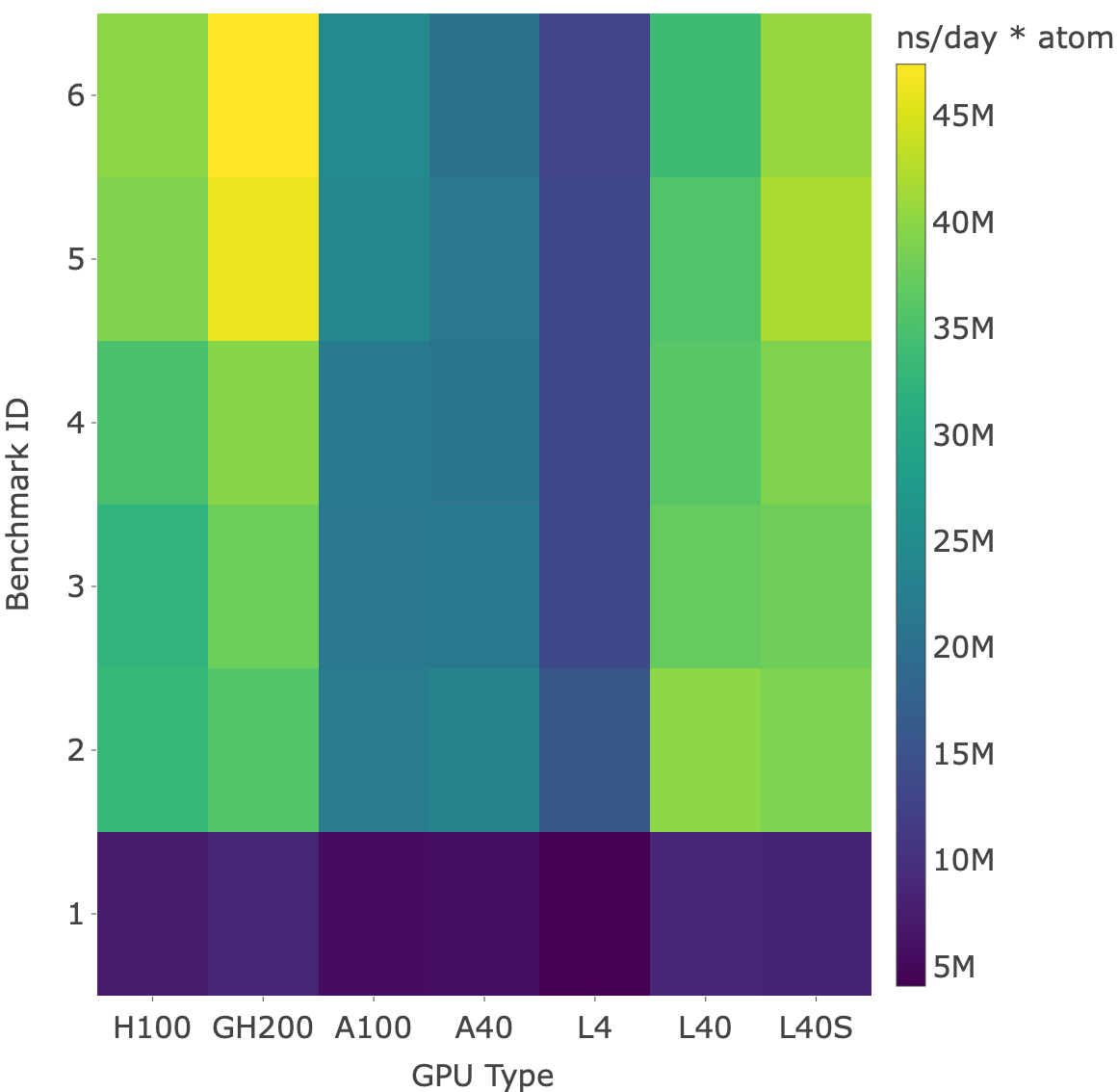}
    }      
    \subfloat[AMBER Performance\label{fig:2md_start0}]{
        \hspace{-0.6em}\includegraphics[width=0.24\textwidth]{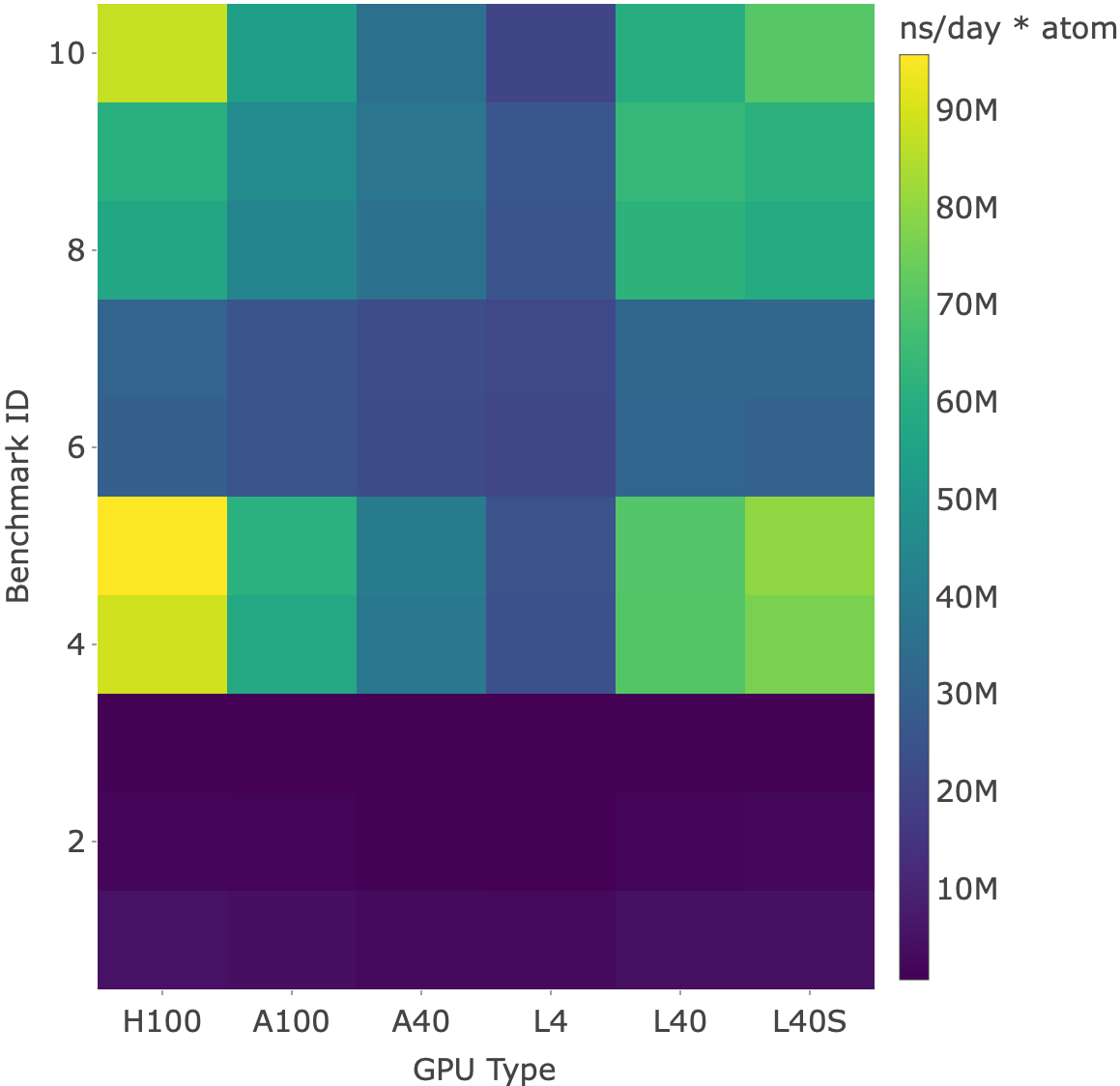}
    }
    
    \subfloat[GROMACS Power\label{fig:rnanvt}]{
         \hspace{-0.4em}\includegraphics[width=0.23\textwidth]{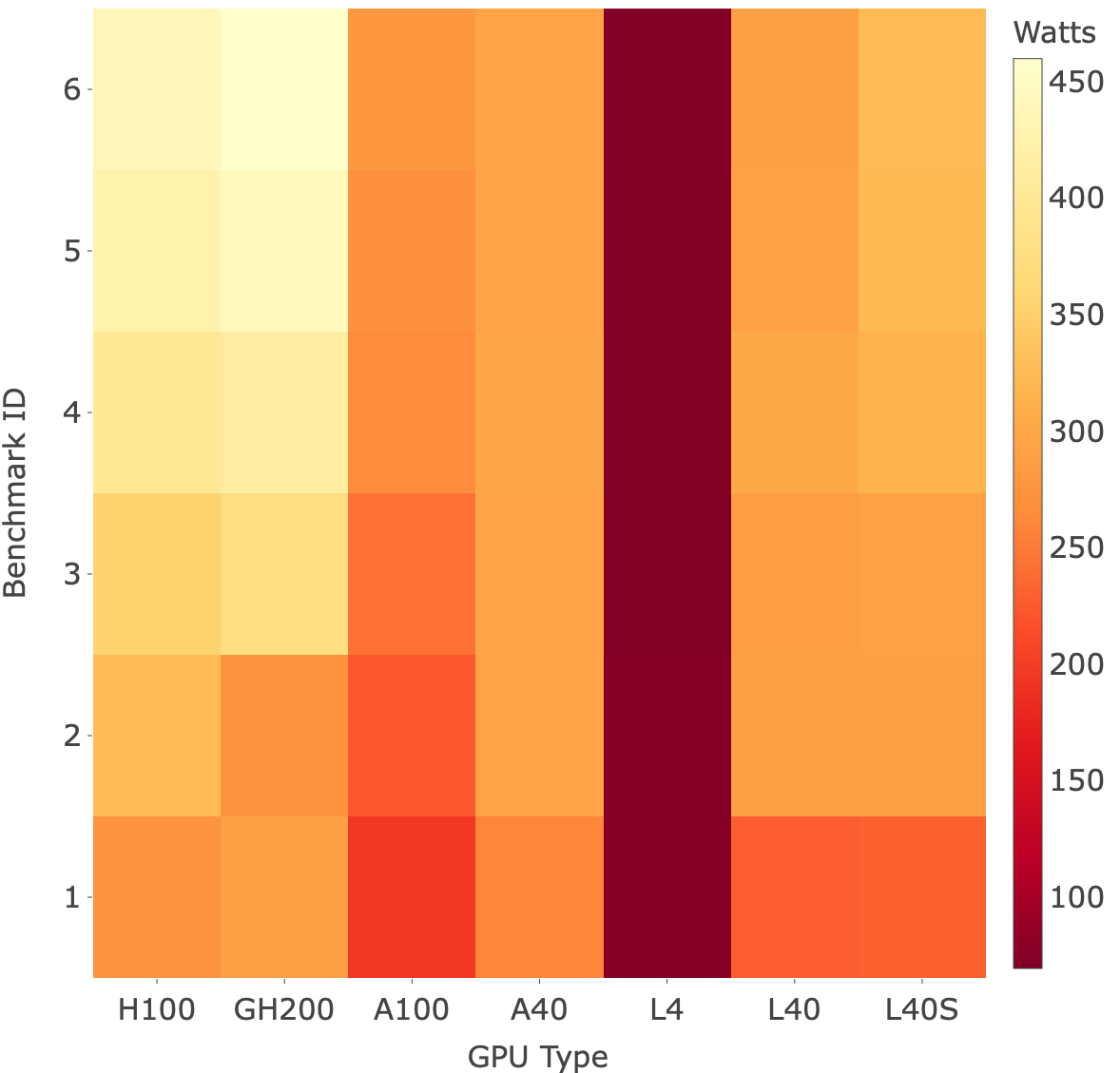}
    }      
    \subfloat[AMBER Power\label{fig:pi_large_test}]{
        \hspace{-0.5em} \includegraphics[width=0.23\textwidth]{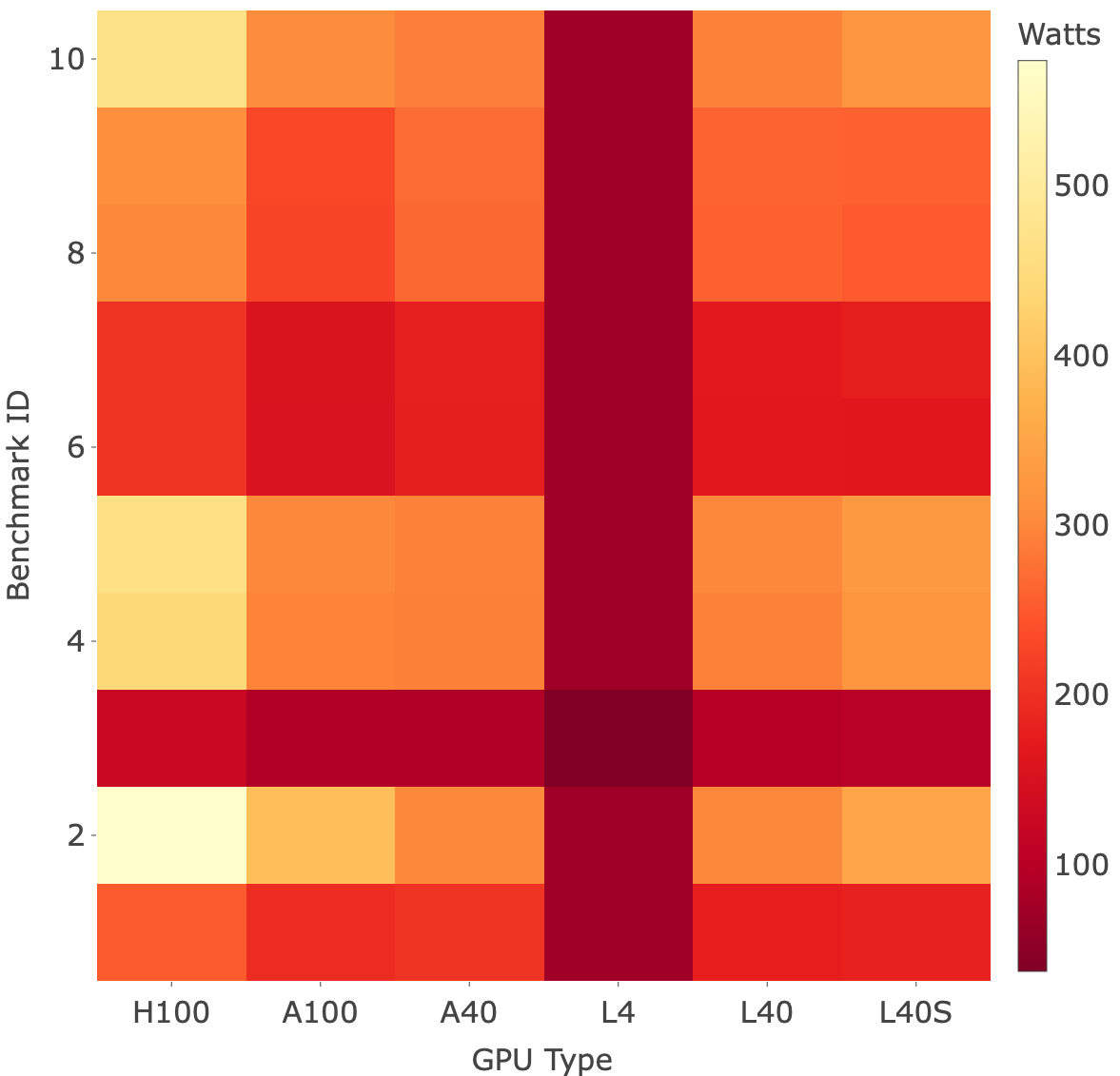}
    }
    \caption{\AAadd{Average} application performance and power draw for GROMACS and AMBER at base frequencies and TDP. Each pair of plots shows performance (top) and corresponding power usage (bottom).}
    \label{fig:apps_perf_power}
\end{figure}

\subsubsection{Performance Modeling}
Wattlytics predicts application throughput on heterogeneous clusters by interpolating benchmark data from representative scientific workloads across coupled frequency domains\AAadd{~\cite{AfzalHW:2025:2}}. It extrapolates beyond the measurable frequency range where necessary.
For a cluster with $n_\text{GPU}$ total accelerators and $g_\text{node}$ GPUs per node, 
aggregate performance for homogeneous setups is:
\begin{equation}\label{eq:eta}
\begin{aligned}
    \tilde{P}_\text{GPU} &= 
        n_\text{GPU} \cdot {P}_\text{GPU}(f_\text{GPU}) \cdot \eta_\text{multi-GPU}
    \\
   \eta_\text{multi-GPU} &= \eta_\text{node}^{\,n_\text{nodes}-1} \cdot \eta_\text{GPU}^{\,n_\text{GPU}-1}
\end{aligned}
\end{equation} 
\AAadd{Here, $\eta_{\text{multi-GPU}} \in (0,1]$ accounts for multi-device efficiency losses arising from inter-node communication and synchronization ($\eta_\text{node}^{\,n_\text{nodes}-1}$) and intra-node contention ($\eta_\text{GPU}^{\,n_\text{GPU}-1}$). Typical ranges for these factors are workload-dependent~\cite{Pall:2020}. The model assumes strong scaling (one workload spanning all GPUs\AAcomm{, a deliberate worst case}); \AAcomm{independent jobs correspond to $\eta_\text{multi-GPU} \approx 1$}. 
As benchmark-specific scalability data is typically unavailable at procurement time, Wattlytics treats scalability as an uncertainty dimension. Extending Wattlytics to allow user-provided scalability curves for strong/weak scaling fitting is left for future work. Hardware-specific scaling via transferable models (e.g., Amdahl-like) could enable compute–communication trade-off estimation without exhaustive benchmarking.
Single-GPU throughput is frequency-dependent:}
\begin{equation}
\begin{aligned}
    {P}_{\text{GPU}}(f_\text{GPU}) &=
    P_{\text{base, GPU}} \cdot \psi_{\text{GPU}}(f_\text{GPU})\\
    \qquad &= \min (P^{\text{max}}_{\text{GPU}}, \beta f_{\text{GPU}} + \gamma), 
\end{aligned}
\end{equation} 
where $P_{\text{base, GPU}}$ is reference throughput at nominal frequency (Figure~\ref{fig:apps_perf_power}), $P^{\text{max}}_{\text{GPU}}$ is the maximum, and $\psi_{\text{GPU}}$ is an empirical frequency-scaling function capturing frequency-dependent performance variations~\cite{AfzalHW:2025:2}.
This allows Wattlytics to evaluate GPU frequency impacts on work-per-TCO and cost-efficiency at device and cluster levels.

\begin{figure*}[t]
    \centering 
    \subfloat[GROMACS - Bench 4]{
        \hspace{-0.6em}\includegraphics[width=0.35\textwidth]{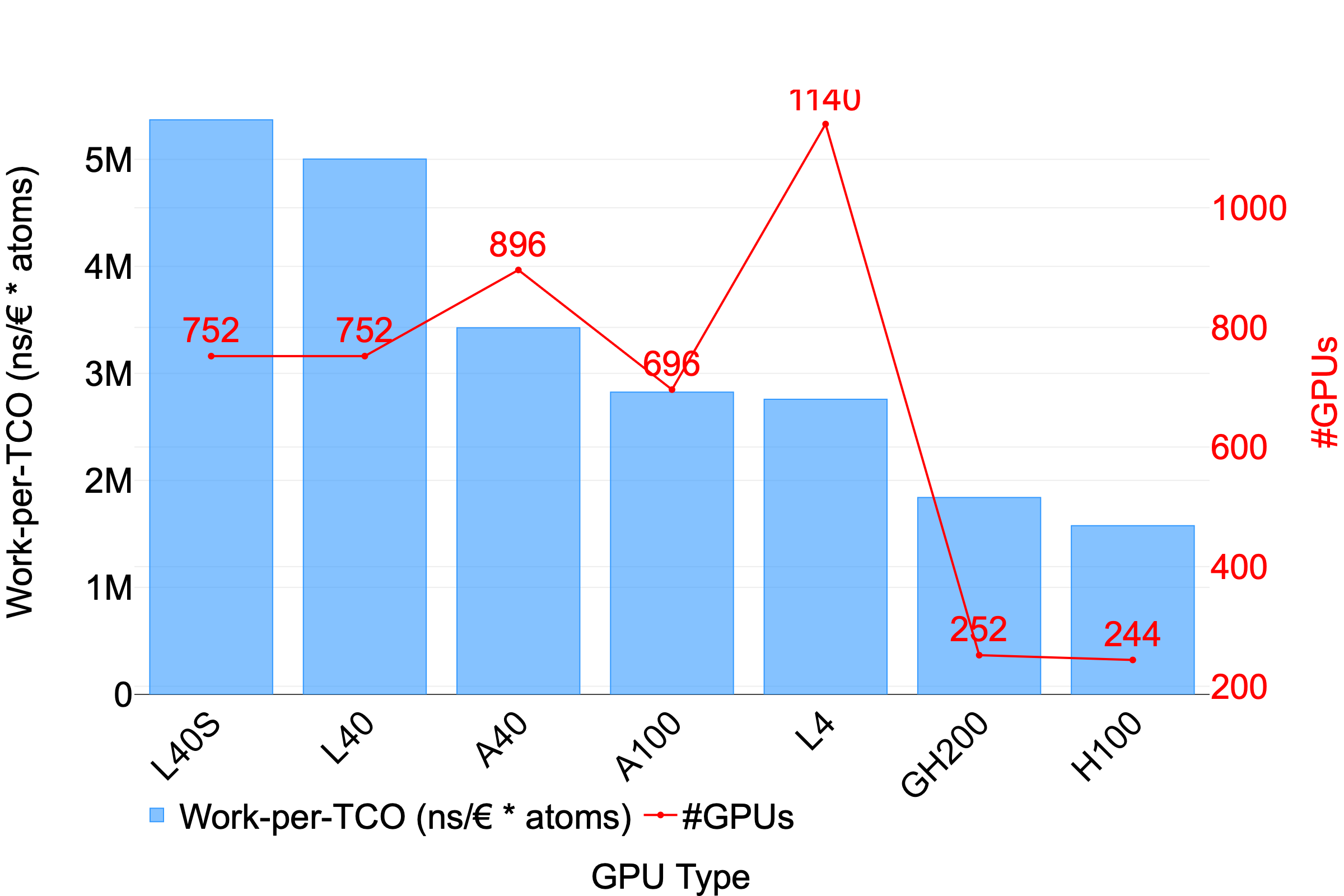}
    } \qquad
    \subfloat[AMBER - Bench 3]{
        \hspace{-0.5em}\includegraphics[width=0.35\textwidth]{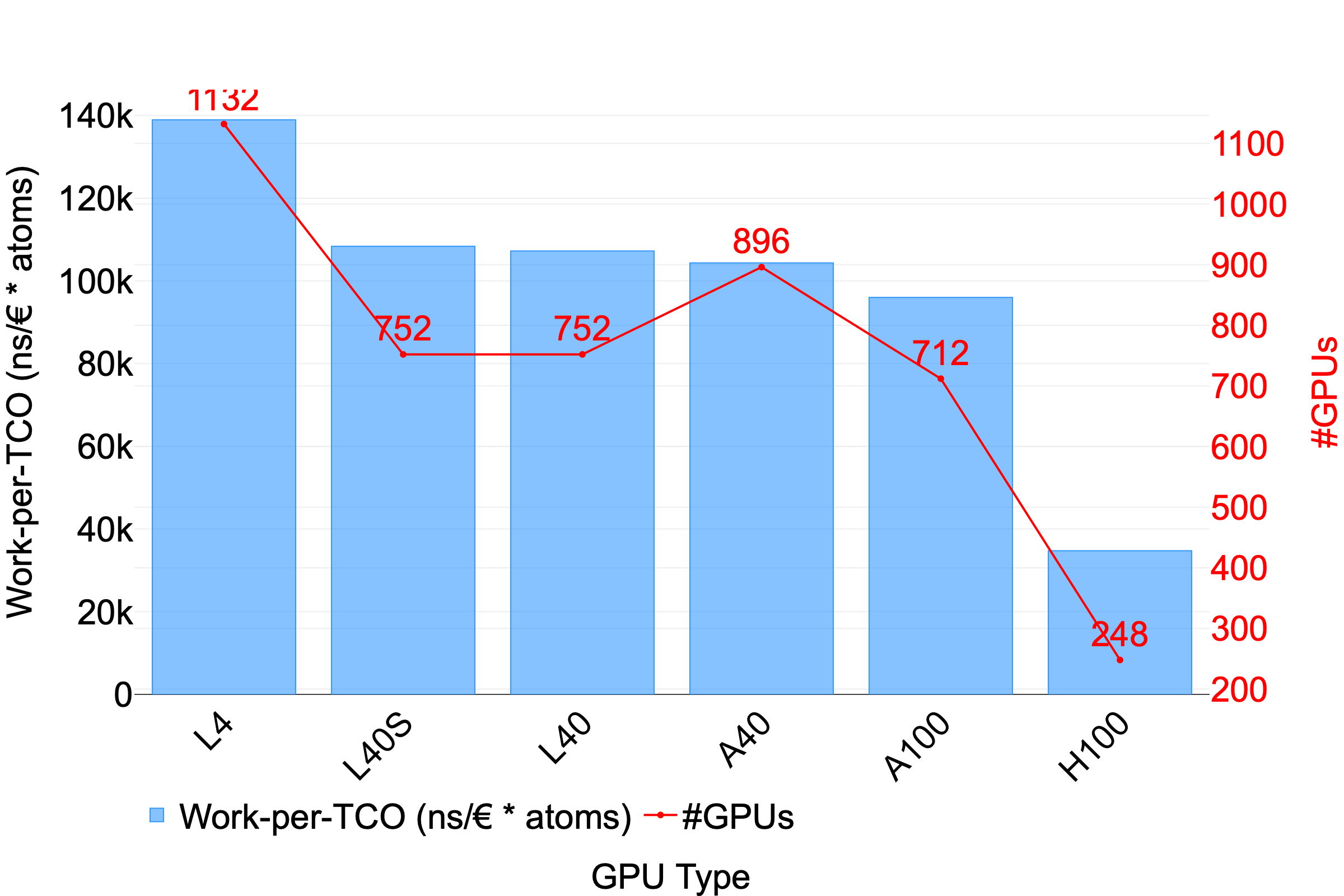}
    } 
    \caption{Work-per-TCO comparison of molecular dynamics workloads in Wattlytics: (a) GROMACS on Benchmark 4, (b) AMBER on Benchmark 3. Experiment links for reproducibility: (a)~\cite{web5}, (b)~\cite{web6}.}
    \label{fig:bench}
\end{figure*}
\subsubsection{Sensitivity and uncertainty modeling}\label{sec:uncertainty}
Wattlytics quantifies how variations in input parameters affect output metrics across heterogeneous GPU configurations using three complementary measures. 
These measures distinguish local slopes (\emph{elasticity}), global variance contributions (\emph{Sobol total-order}), and parameter-specific uncertainty propagation (\emph{Monte Carlo}), supporting interpretable and robust decisions.

\paragraph{Elasticity (local sensitivity)}
For each input parameter $x_i$, Wattlytics computes a discrete, dimensionless elasticity $E_{x_i}$ with respect to the output metric M:
\begin{equation}
E_{x_i} \approx \frac{\partial \mathrm{M}}{\partial x_i} \cdot \frac{x_i}{\mathrm{M}} \times 100\,. 
\end{equation}
The partial derivative $\frac{\partial \mathrm{M}}{\partial x_i}$ can have any sign and is derived analytically from the model equations, avoiding numerical differentiation or finite-difference perturbations. 
Elasticity reflects the local slope of M at the nominal operating point and is valid for small perturbations; values may exceed $\pm100\%$ when changes in $x_i$ induce more-than-proportional responses in M.

\paragraph{Sobol total-order indices (global sensitivity)}
While elasticity captures local sensitivity at a nominal point, Sobol indices measure global sensitivity over finite input ranges (e.g., $\pm20\%$)\AAadd{~\cite{SOBOL:2001}}. 
They quantify the fraction of output variance attributable to each parameter over a finite uncertainty range, including nonlinear effects and interactions with other inputs. Signed elasticity and non-negative Sobol indices measure fundamentally different quantities and only approximately align under linear, independent, and symmetric input assumptions. Wattlytics implements safeguards for zero-valued baseline parameters by replacing zeros with a small value (0.001), preventing misleadingly zero elasticity, zero Monte Carlo standard deviation or meaningless Sobol indices.
Let $A$ and $B$ be independent $N \times n$ sample matrices, where $N=2000$ is the number of  Monte Carlo samples and $n=15$ is the number of input parameters. The total-order Sobol index $S_{T_i}$ for parameter $x_i$ is estimated \AAadd{using the Jansen estimator~\cite{Saltelli:2010} and the ``pick-and-freeze'' method} as
\begin{equation}
S_{T_i} \approx \frac{\frac{1}{N} \sum_{k=1}^{N} \big(\mathrm{M}_A^{(k)} - \mathrm{M}_{A_B^{(i)}}^{(k)}\big)^2}{2 \, \mathrm{Var}(\mathrm{M})} \times 100 \,, 
\end{equation} 
where $\mathrm{M}_A^{(k)} = f(A^{(k)})$, $\mathrm{M}_{A_B^{(i)}}^{(k)} = f(A_B^{(i,k)})$, and $A_B^{(i,k)}$ equals the $k$-th row of $A$ except for the $i$-th parameter, taken from $B$. The numerator captures the squared output difference induced by perturbing $x_i$, while the denominator normalizes by the total output variance, computed using an unbiased estimator.
All parameters are normalized during sampling to avoid numerical scaling issues.

\paragraph{One-at-a-time Monte Carlo (uncertainty propagation)}
Whereas Sobol analysis varies all parameters jointly, this analysis perturbs only $x_i$ within its uncertainty range ($N=2000$ samples, other inputs at baseline) and reports the relative spread.  
The spread of this distribution captures how uncertainty in $x_i$ alone contributes to uncertainty in the output metric.
The relative uncertainty contribution of parameter $x_i$ is defined as
\begin{equation}
U_{x_i} = \frac{\sqrt{\mathrm{Var}\left(\mathrm{M}^{(1)}, \dots, \mathrm{M}^{(N)}\right)}}{\mathrm{M}_0} \times 100\,,
\end{equation}
where $\mathrm{M}_0$ is the baseline output, i.e., the percentage uncertainty in M due to $x_i$ alone, allowing direct comparison of cost, performance, and energy risk drivers across configurations.

\section{Evaluation and Case Studies}\label{sec:evaluation}
Since Wattlytics is designed for decision support, we evaluate it by answering the concrete questions faced by decision makers and HPC operators when selecting, tuning, and operating GPU-based systems. 
We present a sequence of case studies framed as targeted questions (Q1–Q9), each isolating a specific design or operational decision.
These questions examine GPU deployment strategies, frequency and power-cap tuning, TCO composition, parameter sensitivity, and the impact of operational uncertainty.
Our primary optimization objective is to maximize long-term work-per-TCO or, alternatively, minimize power-per-TCO under practical constraints.
Unless stated otherwise, all experiments assume a system lifetime of $T_{\text{life}}=5$ years, a total budget of $B=\euro10\,\text{M}$, multi-GPU efficiency $\eta_\text{node}=0.999$ and $\eta_\text{GPU}=1$ (Q3--Q4: $\eta_\text{node}=0.995$), benchmark 4 of GROMACS, and an electricity price of $C_\text{elec}=\euro0.21/\text{kWh}$.

\subsection{Model accuracy}\label{sec:accuracy}
\AAcomm{End-to-end TCO validation would require observing a cluster over its multi-year lifetime; we therefore validate the two fitted models and bound the unobservable future via uncertainty analysis (Q8--Q9). Power fits match measurements within a few watts ($<$1\%, at most 1.8\% for large H100 benchmarks)~\cite{Afzal:2026sweet}, and throughput fits stay within run-to-run variability ($<$5\%). Propagated to work-per-TCO, the error stays below 6\%, far below the 2.5--4.5$\times$ gaps of Q1; closer gaps are treated as ties. Three further cases were evaluated solely as custom uploads (Section~\ref{sec:extensibility}); we omit their detailed results for brevity and report only deviations from the main results. AMD MI300X power scales linearly with clock, a special case of Eq.~\eqref{eq:piecewise_power}, with errors in the same range (DVFS parameters for NVIDIA and AMD \cite{webLIV}). The synthetic kernels bracket MD: memory-bound STREAM TRIAD barely scales with clock, compute-bound PI-Solver nearly linearly~\cite{AfzalHW:2025:2}. LLM decode (four attention architectures on H200, an AI case~\cite{Ma:2026}) draws only 137--300\,W of its 700\,W TDP, so TDP-based estimates would overstate its power 2--5$\times$; recommendations thus need per-workload profiles. Otherwise, results agree with the main findings. Results below cover MD on NVIDIA GPUs with one workload at utilization $U_\text{sys}$, worst-case parallel efficiency (Eq.~\eqref{eq:eta}), single-center prices, and independent uniform uncertainties: upshots 1, 2, 5, 6, and 9 (below) hold for capacity workloads at realistic prices; 3 and 4 depend on the worst-case efficiency model.}

\subsection*{Q1: Which GPU maximizes work-per-TCO with fixed-budget? (reproducibility artifacts:~\cite{web5})}
\label{sec:fixed_budget}
We begin with the most common procurement question: Given a fixed capital budget, which GPU configuration maximizes long-term scientific output?
For GROMACS, although GH200 and H100 deliver the highest single-device throughput, their high acquisition and power costs limit cluster scale.
In contrast, lower-cost GPUs such as L4 and L40S (Table \ref{tab:gpu_specs_perf}) can be deployed at substantially higher multiplicity within the same budget.
Relative to GH200, L4 delivers 2.0--3.7$\times$ lower performance with 4.0--6.5$\times$ lower power, while L40S achieves near-parity performance (0.9--1.2$\times$ lower) with 0.9--1.4$\times$ lower power draw (Figure \ref{fig:apps_perf_power}).
As a result, scaling out L4 or L40S GPUs yields \emph{3.2--3.8$\times$ higher aggregate work-per-TCO} for 6 benchmarks over five years (Figure \ref{fig:bench}(a))\AAcomm{\footnote{All figure fonts will be enlarged in the camera-ready version; interactive full-resolution versions are available via \cite{web1}--\cite{web11}.}, a lower bound, as $\eta_\text{node}=0.999$ already costs L4 25\% of ideal throughput versus 6\% for H100}.
\begin{figure*}[t]
    \centering       
        \includegraphics[width=0.6\textwidth]{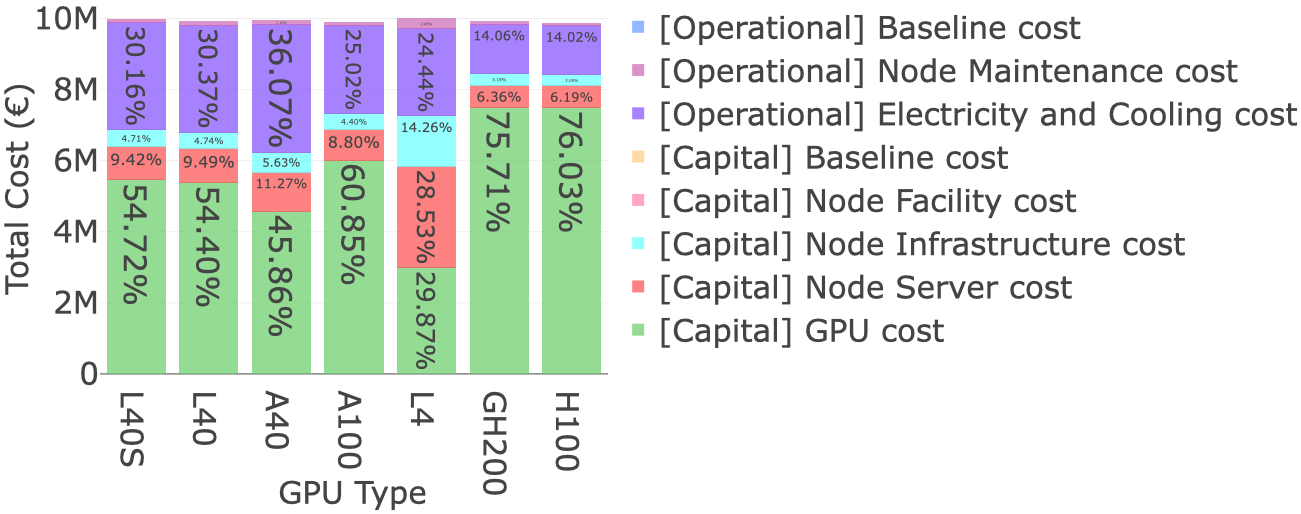}
    \caption{TCO breakdown under a fixed $\euro10\,\text{M}$ budget for GROMACS on Benchmark 4 in Wattlytics; see~\cite{web5} experiment link for reproducibility.}
    \label{fig:tcobreakdown}
\end{figure*}
Despite superior per-device energy efficiency, scale-out deployments incur higher \emph{aggregate} electricity and cooling costs due to increased GPU counts: total energy and cooling costs are approximately 1.4$\times$ higher for L4 and 2$\times$ higher for L40S compared to GH200/H100 systems.
Moreover, L4's cost-effectiveness is partially offset by higher server and infrastructure overheads, which account for roughly 50\% of its capital share, compared to only 15\% for L40S (Figure~\ref{fig:tcobreakdown}).
Overall, L40S provides the best work-per-TCO for GROMACS workloads.
To assess workload sensitivity, we repeat the analysis for AMBER across eleven benchmarks. Relative to H100, L4 delivers 1.1--4.7$\times$ lower performance with 2.9--8.0$\times$ power savings, while L40S delivers 0.9--1.3$\times$ lower performance with 1.2--1.6$\times$ power savings.
Despite lower raw throughput, deploying more L4 or L40S GPUs within the same budget yields \emph{2.5--4.5$\times$ higher lifetime work-per-TCO}, confirming that optimal GPU choices are workload-dependent but consistently favor energy-efficient scaling under budget constraints.
\vspace{2pt}\highlight{\emph{Upshot 1}:
Wattlytics exposes non-obvious, multidimensional trade-offs in GPU deployment, demonstrating that \emph{budget-aware multi-GPU scaling} often dominates peak single-device performance in determining long-term scientific output.
}
\begin{figure*}[t]
    \centering       
    \subfloat[Fixed-budget (\euro10\,\text{M})]{
        \includegraphics[width=0.35\textwidth]{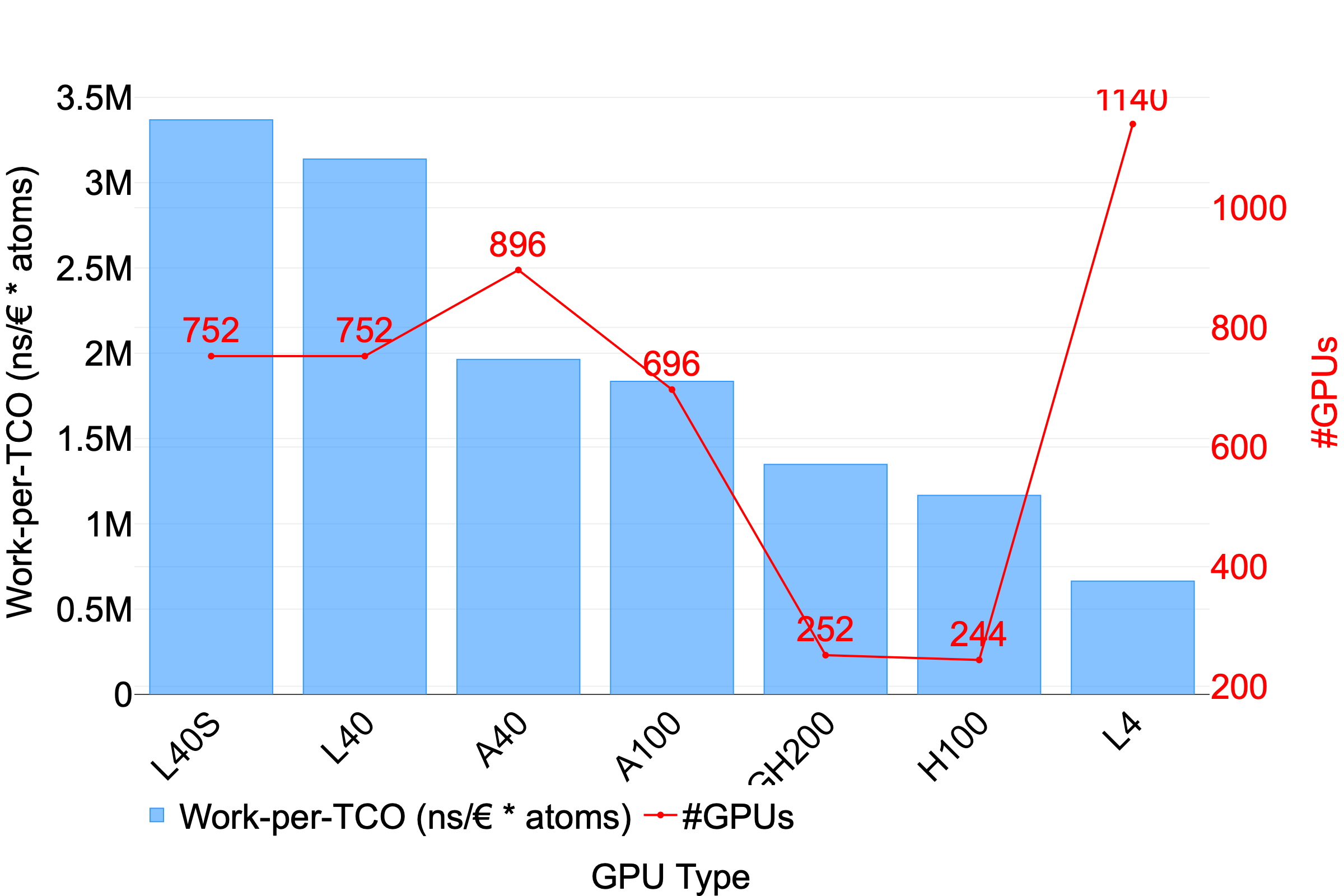}
    } \quad
    \subfloat[Fixed-power (78 kW)]{
        \includegraphics[width=0.35\textwidth]{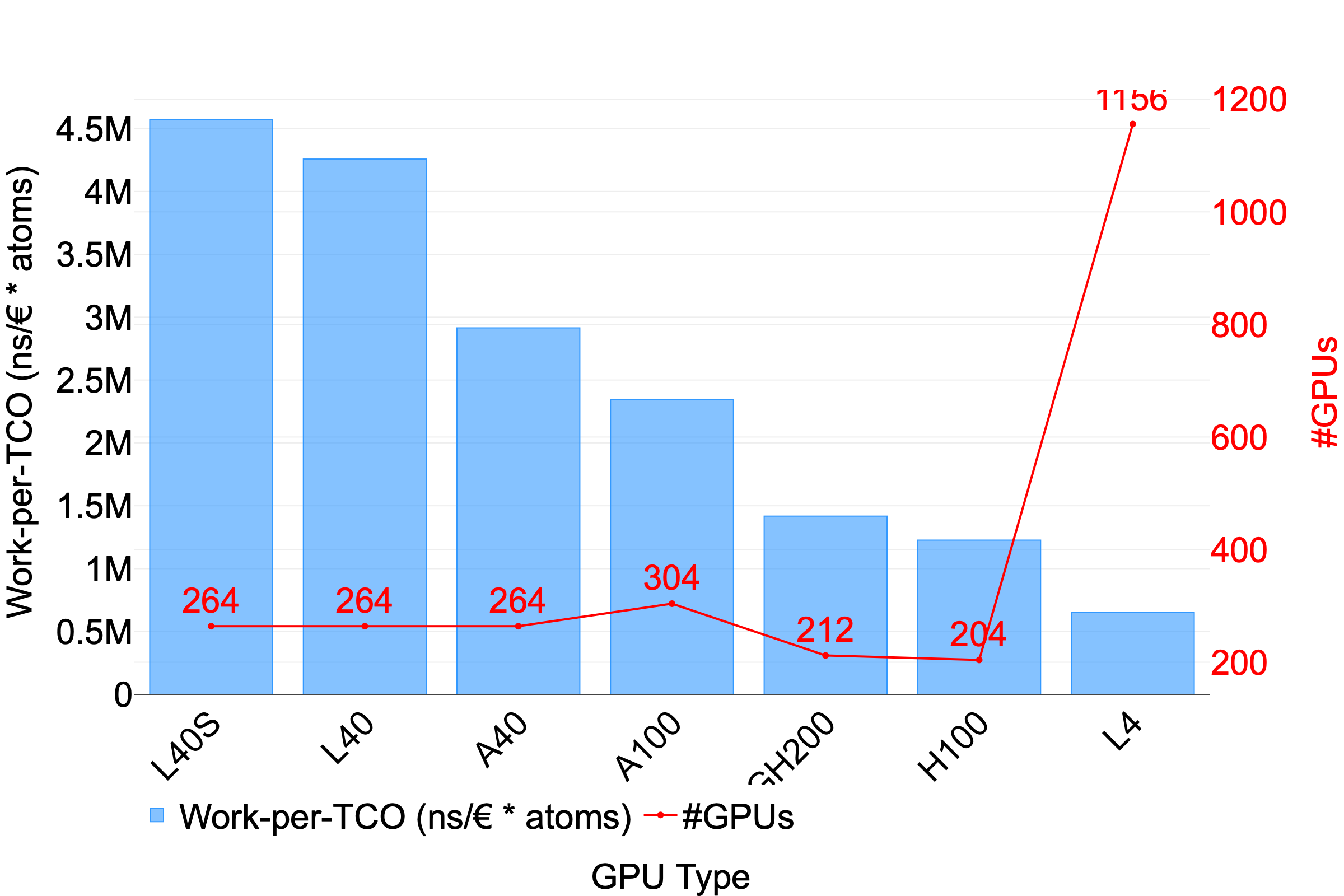}
    }

    \subfloat[Fixed-perf. ($8.9 \times 10^9$ ns/day*atom)]{
        \includegraphics[width=0.35\textwidth]{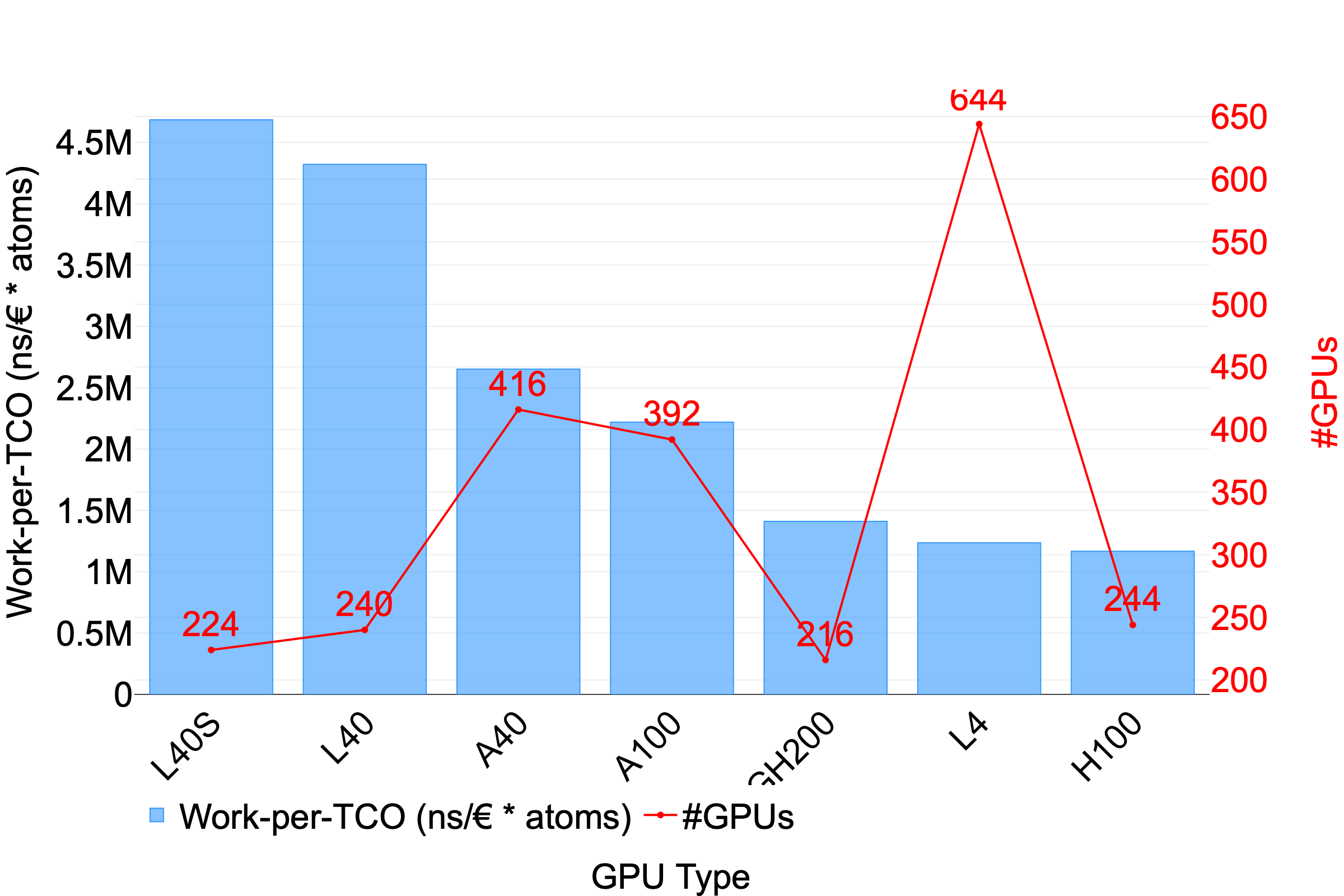}
    }\quad
    \subfloat[Fixed-GPU count (248)]{
        \includegraphics[width=0.35\textwidth]{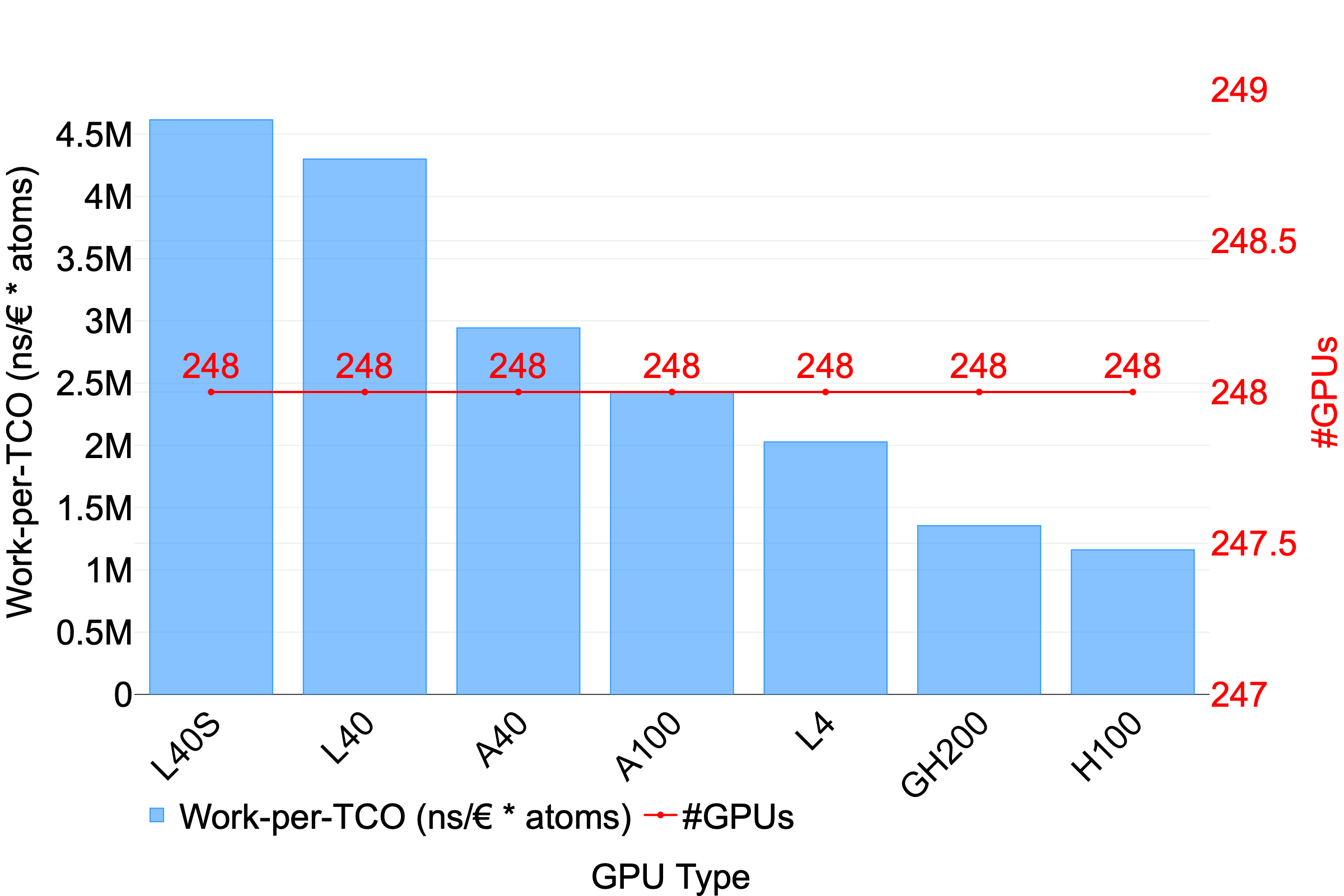}
    } 
    \caption{Comparison of deployment strategies under different constraints in Wattlytics with a budget cap of $B = \euro10\,\text{M}$ and multi-GPU efficiency $\eta_\text{node}=0.995$ ($\eta_\text{GPU}=1$) for GROMACS on Benchmark 4. Experiment links for reproducibility: (a)~\cite{web1}, (b)~\cite{web2}, (c)~\cite{web3}, (d)~\cite{web4}.}
    \label{fig:deployment}
\end{figure*}

\subsection*{Q2: When do lowest-power GPUs become most cost effective? (reproducibility artifacts:~\cite{web6})}
While high-end GPUs typically dominate in raw throughput, we observe notable inversions under AMBER Benchmark~3, which exhibits low thermal intensity and modest performance demands, making the L4 the most cost-effective GPU despite its lower peak performance (Figure~\ref{fig:bench}(b)). 
Its low power draw enables greater deployment scale and superior work-per-TCO under a fixed budget.
This inversion does not occur for GROMACS benchmarks or other AMBER benchmarks, which favor mid-range GPUs such as L40S for superior work-per-TCO, underscoring that \emph{optimal GPU selection is workload-dependent}.
\vspace{2pt}\highlight{\emph{Upshot 2}:
Wattlytics identifies workload-specific optimization, revealing regimes in which slower, energy-efficient GPUs outperform higher-end accelerators under budget or energy constraints.  
}

\subsection*{Q3: Is it better to run one job per GPU or parallelize across multiple GPUs? (reproducibility artifacts:~\cite{web1},~\cite{web5})}
Multi-GPU execution incurs efficiency losses due to communication and synchronization overheads, captured by $\eta_{\text{multi-GPU}}$.
This raises a key operational question: should workloads be executed as independent single-GPU jobs or parallelized across multiple GPUs?
Using Wattlytics, we evaluate this trade-off by reducing $\eta_\text{node}$ from 0.999 (Figure~\ref{fig:bench}(a)) to 0.995 (Figure~\ref{fig:deployment}(a)) and comparing the resulting work-per-TCO.
We observe that even small per-GPU efficiency losses are sufficient to reverse earlier conclusions: under reduced multi-GPU efficiency, lower-performing GPUs such as L4 become less cost-effective than high-end GPUs (GH200/H100), despite their lower power consumption. \AAcomm{Since Eq.~\eqref{eq:eta} compounds over nodes, 0.995 is a pessimistic capability scenario (1149 L4s retain 24\% of ideal throughput vs.\ 75\% at 0.999); capacity workloads ($\eta\approx1$) keep Q1's ranking.}  
\vspace{2pt}\highlight{\emph{Upshot 3}:
Wattlytics reveals that small multi-GPU efficiency losses can alter optimal deployment strategies, shifting the advantage from scale-out, low-power GPUs to fewer, higher-performance accelerators \AAcomm{for tightly coupled jobs}.}

\subsection*{Q4: Do optimal choices persist under alternative constraints? (reproducibility artifacts:~\cite{web1} to \cite{web4})}
In practice, deployments are constrained not only by capital budgets but also by performance targets, power or cooling limits, and fixed GPU counts.
Across all cases (Figure~\ref{fig:deployment}(a)--(d)), we enforce a global budget constraint of $\euro10\,\text{M}$ and account for non-ideal multi-GPU efficiency ($\eta_{\text{multi-GPU}}<1$), thereby incorporating realistic scaling overheads.
Although absolute rankings shift across constraint modes, the qualitative behavior remains stable: Optimal configurations are those that best balance performance, power, cost, and scaling efficiency within the fixed budget.
When performance targets are imposed (Figure~\ref{fig:deployment}(b)), high-end GPUs such as GH200 and H100 are not consistently favored due to their high acquisition costs, despite achieving required throughput with fewer devices and lower cumulative efficiency loss.
Under power constraints (Figure~\ref{fig:deployment}(c)), energy-efficient GPUs such as L4 likewise fail to dominate, as their scale-out advantage is offset by compounding multi-GPU efficiency losses at large deployment sizes.
Similarly, under fixed GPU counts (Figure~\ref{fig:deployment}(d)), higher-cost GPUs consume a disproportionate share of the budget, while lower-cost GPUs incur greater aggregate efficiency penalties, preventing either extreme from consistently maximizing work-per-TCO.
\vspace{2pt}\highlight{\emph{Upshot 4}:
Even under additional deployment constraints, the fixed budget and non-ideal multi-GPU efficiency govern system-level outcomes, and Wattlytics captures performance--power--cost interactions including non-ideal scaling.}

\subsection*{Q5: How do system-level design choices affect GPU rankings? (reproducibility artifacts:~\cite{web7})}
We evaluate GPU density by scaling L4 deployments from 4 to 8 GPUs per node.
At 4 GPUs per node, L4 underperforms A40/A100 due to higher per-node overheads.
Increasing density to 8 GPUs per node improves amortization of server and infrastructure costs, reversing the ranking and enabling L4 to outperform A40/A100 in work-per-TCO.
This demonstrates that optimal GPU selection depends not only on device performance but also on system-level integration and overheads.
\vspace{2pt}\highlight{\emph{Upshot 5}: GPU rankings are sensitive to node-level design choices; increasing GPU density can significantly improve cost-effectiveness for energy-efficient devices.}

\subsection*{Q6: How sensitive are GPU rankings to system lifetime? (reproducibility artifacts:~\cite{web8})}
We vary system lifetime from 1 to 9 years to examine the relative influence of capital versus operational costs.
Short lifetimes amplify capital expenditure, favoring high-performance GPUs, whereas longer lifetimes increase the impact of energy costs, shifting optimal rankings toward lower-power GPUs.
For typical HPC lifetimes (3 to 7 years), rankings remain stable.
\vspace{2pt}\highlight{\emph{Upshot 6}: Optimal GPU selection is largely insensitive to typical HPC lifetimes; cost-efficiency rankings remain stable under realistic operational horizons.}

\begin{figure*}[t]
    \centering
    \subfloat[What-if ``energy price'' scenario]{
        \hspace{-2.8em}\includegraphics[width=0.298\textwidth]{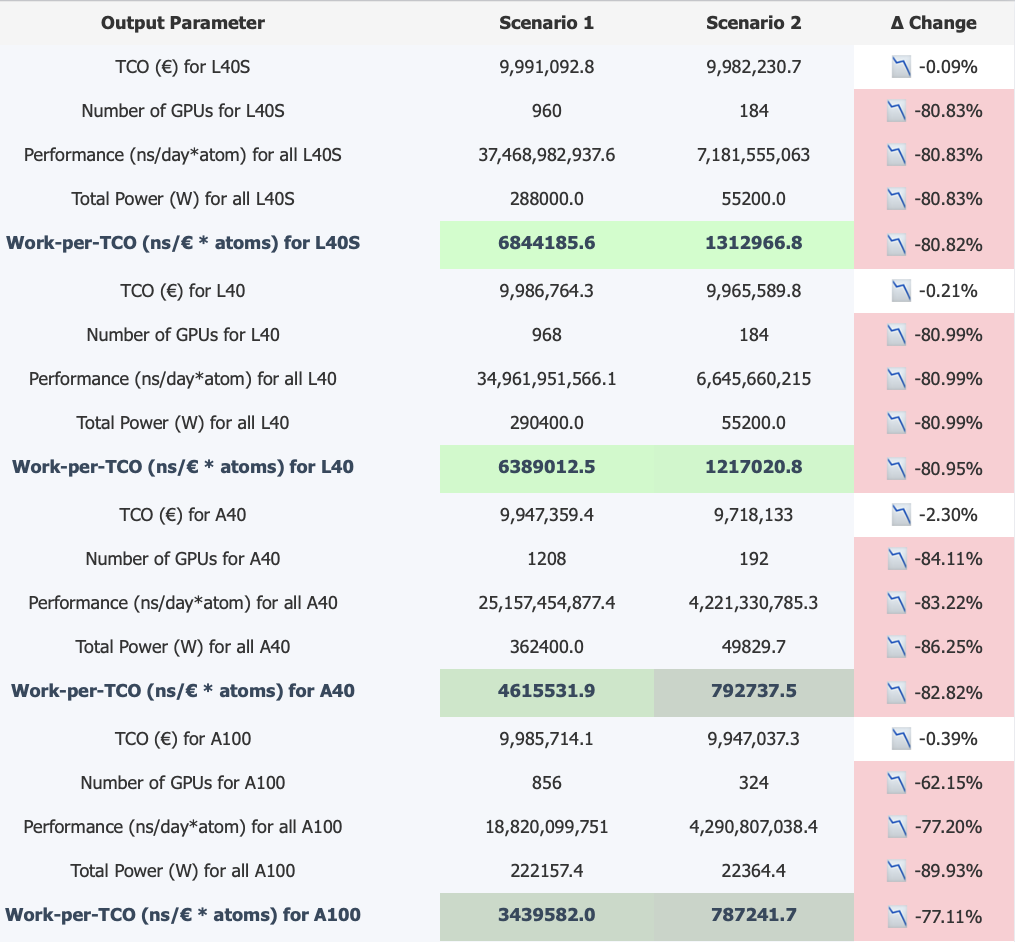}
    }
    \subfloat[Sensitivity and uncertainty analysis]{
    \includegraphics[scale=0.236]{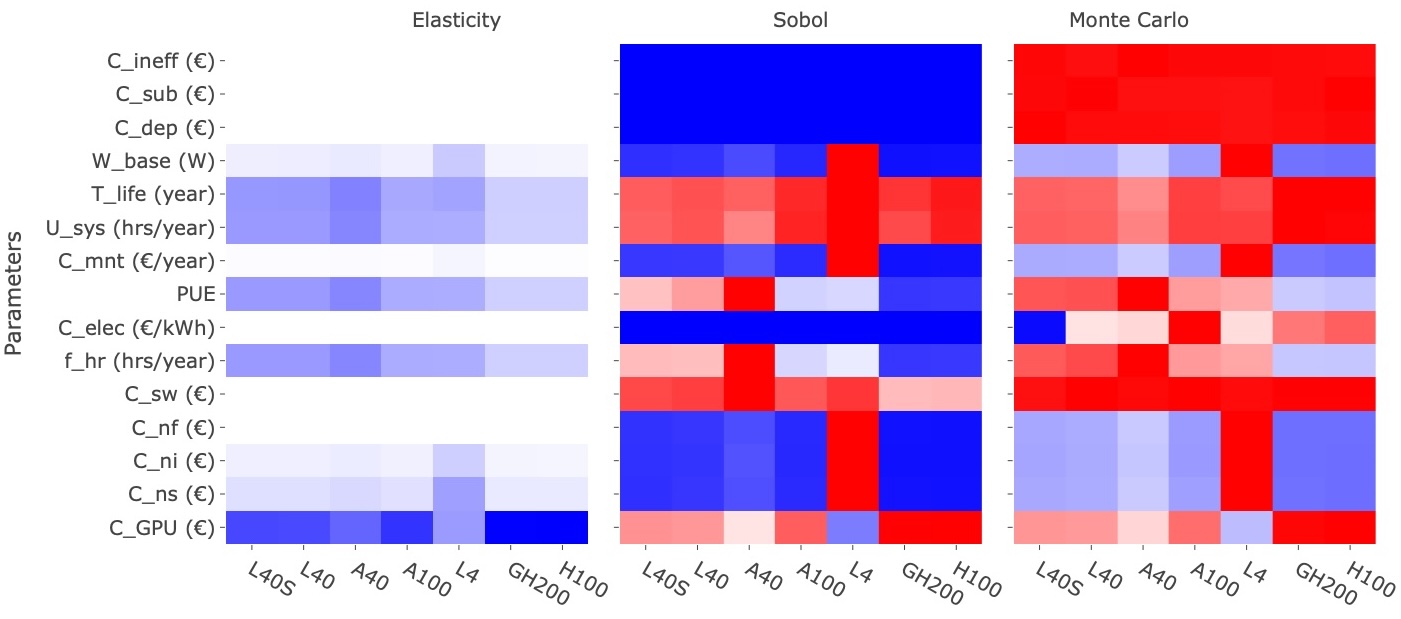}\put(0,11){\includegraphics[height=0.235\textwidth]{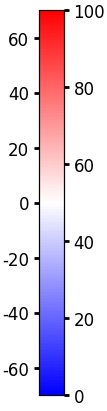}}
    }
    \caption{(a) What-if electricity price scenario~\cite{web11}; (b) Cross-GPU heatmaps comparing elasticity, Sobol indices, and Monte Carlo results, showing the relative impact of parameters on work-per-TCO. Blue and red colors represent 0\% and 100\% for Sobol and Monte Carlo indices, while they correspond to -max and +max for Elasticity; see reproducible experiment link at ~\cite{web10}, which additionally shows the bar charts displaying per-GPU contributions.}
    \label{fig:SensitivityA}
\end{figure*}
\subsection*{Q7: Can operational tuning boost work-per-TCO? (reproducibility artifacts:~\cite{web9})}
Hardware choice is not the only lever for efficiency. 
Using a GROMACS workload on A100 GPUs, we progressively reduce the GPU graphics clock $f_{\text{GPU}}$ from 2.04\,GHz (extrapolated via the fitted models beyond the 1.41\,GHz vendor maximum) to 1.2\,GHz and predict performance, power, and work-per-TCO. 
Moderate frequency reductions yield substantial energy savings with minimal performance loss, improving both work-per-watt and work-per-TCO, whereas aggressive underclocking causes nonlinear performance degradation that offsets energy gains. 
Wattlytics identifies workload-specific “knee points” where further frequency adjustments become counterproductive, enabling operators to optimize the trade-off between energy efficiency and sustained performance. 
Such extrapolations let users evaluate hypothetical operating points.
\vspace{2pt}\highlight{\emph{Upshot 7}: Moderate GPU frequency reduction can reduce power and energy costs with minimal throughput loss, improving work-per-watt-per-TCO and recovering some benefits of hardware replacement at zero capital expense.}

\subsection*{Q8: Which parameters drive cost, and which drive risk? (reproducibility artifacts:~\cite{web10})}
We perform systematic sensitivity analyses using elasticity, Sobol, and Monte Carlo methods to quantify the impact of input parameters on work-per-TCO and power-per-TCO; see Figure \ref{fig:SensitivityA}(b).
GPU hardware cost dominates work-per-TCO, where $E_\text{H100\_cost}=-76\%$ indicates that a 1\% increase in H100 cost yields a 0.76\% decrease in work-per-TCO. For high-capital GPUs like H100, GPU cost exhibits the highest elasticity, indicating that total cost is capital-dominated. Also, its uncertainty has the highest impact on variability: perturbing the H100 price by $\pm20\%$ yields the maximum normalized indices ($S_\text{H100\_cost}\approx U_\text{H100\_cost}\approx100\%$). In contrast, node maintenance for H100 contributes minimally to the mean work-per-TCO; its $\pm20\%$ variation yields a normalized Sobol index of only 3.5\% ($S_\text{H100\_node maintenance}=3.5\%, U_\text{H100\_node maintenance}=22\%$). System usage has moderate elasticity ($-14\%$) but high normalized indices ($S_\text{H100\_System usage}=90\%, U_\text{H100\_System usage}\approx97\%$). Thus, GPU cost sets the baseline, whereas system usage drives financial risk. Low-power GPUs such as L4 are most sensitive to fixed node costs (infrastructure, server, maintenance), with Sobol and Monte Carlo indices near 100\%, highlighting their vulnerability to infrastructure uncertainties. In contrast, operational-cost-heavy GPUs like A40 show the greatest work-per-TCO volatility from uncertainties in software, electricity, PUE, and system usage.    
\vspace{2pt}\highlight{\emph{Upshot 8}: Cost drivers (GPU hardware cost) dominate the baseline work-per-TCO, while risk drivers (operational or revenue uncertainties) can shift GPU rankings. Wattlytics helps users distinguish between these, turning HPC design into a well-informed, explainable decision process.}

\subsection*{Q9: Resilience to energy price volatility (reproducibility artifacts:~\cite{web11})} 
We evaluate electricity price sensitivity by sampling $C_{\text{elec}} \sim \mathcal{U}(0.06, 2.36)\,\euro/\text{kWh}$; we treat 0.06--0.40\,\euro/kWh as the realistic range and higher values, reached only at energy-crisis peaks, as stress tests.
Rising energy prices elevate capital-dominated GH200 GPUs in work-per-TCO rankings, allowing them to outperform A40. Low-power GPUs like L4 are more sensitive due to their larger relative energy share, reducing effective work-per-TCO, while L40S remains most cost-effective, achieving 2--4$\times$ higher work-per-TCO than high-end GPUs (see a snippet of the what-if scenario in Figure~\ref{fig:SensitivityA}(a))\AAadd{; ranking changes occur in extreme stress-test scenarios.} Higher system-level PUE further amplifies electricity and cooling costs, disproportionately impacting energy-intensive deployments. Wattlytics thus captures differential resilience of GPU configurations under volatile operational costs.  
\vspace{2pt}\highlight{\emph{Upshot 9}: GPUs with larger relative energy consumption are more affected by electricity price fluctuations. Wattlytics enables operators to identify configurations that allow favorable work-per-TCO under volatile energy and cooling costs.}

\section{Conclusion and Future Work} \label{sec:conclusion}
Wattlytics unifies performance, power, and cost modeling in an interactive, scenario-driven platform for systematic exploration of GPU-based HPC systems. Unlike conventional profilers or TCO calculators, it captures non-obvious, multi-dimensional trade-offs, including budget-aware scaling, workload-specific GPU selection, and multi-GPU efficiency losses, while accounting for operational variability and uncertainty. Our case studies across GROMACS and AMBER demonstrate that optimal GPU deployment often favors energy-efficient scale-out strategies, though small efficiency drops or budget constraints can shift the advantage to high-performance accelerators. Wattlytics enables quantitative evaluation of design choices under budget, power, performance, and TCO constraints. It highlights the sensitivity of cost and risk to GPU hardware, node infrastructure, and electricity prices, supporting informed and robust decision-making. 
\AAcomm{Under the modeled assumptions, it} identifies configurations that \AAcomm{may} improve sustainability without sacrificing scientific output.
Its decision-oriented evaluation demonstrates that optimal GPU choices depend jointly on workload, system design, operational tuning, and uncertainty, dimensions that cannot be captured by performance- or cost-only models.
By combining transparent models, efficiency metrics, and interactive analytics, Wattlytics empowers HPC system designers and operators to maximize long-term work-per-TCO while maintaining energy efficiency and operational robustness.

\paragraph*{Future Work}
\AAadd{Wattlytics is intentionally application- and hardware-agnostic, designed for forward compatibility. Using the \emph{custom upload} option, it applies to any ``cooler'' and ``hotter'' workload (including multi-phase codes such as climate modeling and AI training such as MLPerf benchmarks to assess Tensor Core utilization vs. FP32 throughput) that exhibits measurable frequency scaling, generating distinct \emph{work-per-TCO} signatures.}
Future work will 
extend preloaded profiles to further emerging GPU architectures, mixed-node AI/HPC workloads, CPU uncore/core frequency tuning, and REST-based APIs for automated scenario analysis and scheduler integration (e.g., Slurm energy plugins~\cite{slurmenergyplugin}). 
\AAadd{More generally, modeling HPC workloads as mixes of concurrent jobs with distinct scaling behaviors and sizes, where system throughput is expressed as an aggregate over application fractions with individual efficiencies, is also planned.} 

\section*{Acknowledgment}
The authors gratefully acknowledge the HPC resources provided by the Erlangen National High Performance Computing Center (NHR@FAU) at FAU Erlangen-Nürnberg.
NHR funding is provided by the German Federal Ministry of Education and Research and the state governments participating on the basis of the resolutions of the GWK for the national high-performance computing at universities (www.nhr-verein.de/unsere-partner) by federal and Bavarian state authorities. 
NHR@FAU hardware is partially funded by the German Research Foundation (DFG) -- 440719683.

\bibliographystyle{IEEEtran}  
\bibliography{references} 

\begin{thebibliography}{10}
\providecommand{\url}[1]{#1}
\csname url@samestyle\endcsname
\providecommand{\newblock}{\relax}
\providecommand{\bibinfo}[2]{#2}
\providecommand{\BIBentrySTDinterwordspacing}{\spaceskip=0pt\relax}
\providecommand{\BIBentryALTinterwordstretchfactor}{4}
\providecommand{\BIBentryALTinterwordspacing}{\spaceskip=\fontdimen2\font plus
\BIBentryALTinterwordstretchfactor\fontdimen3\font minus \fontdimen4\font\relax}
\providecommand{\BIBforeignlanguage}[2]{{%
\expandafter\ifx\csname l@#1\endcsname\relax
\typeout{** WARNING: IEEEtran.bst: No hyphenation pattern has been}%
\typeout{** loaded for the language `#1'. Using the pattern for}%
\typeout{** the default language instead.}%
\else
\language=\csname l@#1\endcsname
\fi
#2}}
\providecommand{\BIBdecl}{\relax}
\BIBdecl

\bibitem{green500}
J.~Dongarra, H.~Meuer, H.~Simon, M.~Meuer, and E.~Strohmaier, ``{The} {Green500} {List}: {Energy-Efficient} {Supercomputers},'' \url{https://www.top500.org/lists/green500}, Nov 2025.

\bibitem{SHAO:2022}
\BIBentryALTinterwordspacing
X.~Shao, Z.~Zhang, P.~Song, Y.~Feng, and X.~Wang, ``{A review of energy efficiency evaluation metrics for data centers},'' \emph{Energy and Buildings}, vol. 271, p. 112308, 2022. [Online]. Available: \url{https://doi.org/10.1016/j.enbuild.2022.112308}
\BIBentrySTDinterwordspacing

\bibitem{Klemick:2019}
\BIBentryALTinterwordspacing
H.~Klemick, E.~Mansur, D.~Raimi, and J.~Shapiro, ``{How do data centers make energy efficiency investment decisions? Qualitative evidence from focus groups and interviews},'' \emph{Energy Efficiency}, vol.~12, no.~5, pp. 1359--1377, 2019. [Online]. Available: \url{https://doi.org/10.1007/s12053-019-09782-2}
\BIBentrySTDinterwordspacing

\bibitem{Kutzner:2015}
\BIBentryALTinterwordspacing
C.~Kutzner, S.~P{\'a}ll, M.~Fechner, A.~Esztermann, B.~L. de~Groot, and H.~Grubm{\"u}ller, ``Best bang for your buck: {GPU} nodes for {GROMACS} biomolecular simulations,'' \emph{Journal of Computational Chemistry}, vol.~36, no.~26, pp. 1990--2008, 2015. [Online]. Available: \url{https://doi.org/10.1002/jcc.24030}
\BIBentrySTDinterwordspacing

\bibitem{Kutzner:2019}
\BIBentryALTinterwordspacing
------, ``More bang for your buck: Improved use of {GPU} nodes for {GROMACS} 2018,'' \emph{Journal of Computational Chemistry}, vol.~40, no.~27, pp. 2418--2431, 2019. [Online]. Available: \url{https://doi.org/10.1002/jcc.26011}
\BIBentrySTDinterwordspacing

\bibitem{Fan:2020}
\BIBentryALTinterwordspacing
K.~Fan, B.~Cosenza, and B.~Juurlink, ``{Accurate Energy and Performance Prediction for Frequency-Scaled GPU Kernels},'' \emph{Computation}, vol.~8, no.~2, 2020. [Online]. Available: \url{https://doi.org/10.3390/computation8020037}
\BIBentrySTDinterwordspacing

\bibitem{Guerreiro:2018}
\BIBentryALTinterwordspacing
J.~Guerreiro, A.~Ilic, N.~Roma, and P.~Tomas, ``{GPGPU Power Modeling for Multi-domain Voltage-Frequency Scaling},'' in \emph{2018 IEEE International Symposium on High Performance Computer Architecture (HPCA)}.\hskip 1em plus 0.5em minus 0.4em\relax Los Alamitos, CA, USA: IEEE Computer Society, 2018, pp. 789--800. [Online]. Available: \url{https://doi.org/10.1109/HPCA.2018.00072}
\BIBentrySTDinterwordspacing

\bibitem{MEI:2017}
\BIBentryALTinterwordspacing
X.~Mei, Q.~Wang, and X.~Chu, ``{A survey and measurement study of GPU DVFS on energy conservation},'' \emph{{Digital Communications and Networks}}, vol.~3, no.~2, pp. 89--100, 2017. [Online]. Available: \url{https://doi.org/10.1016/j.dcan.2016.10.001}
\BIBentrySTDinterwordspacing

\bibitem{accelwattch}
\BIBentryALTinterwordspacing
V.~Kandiah, S.~Peverelle, M.~Khairy, J.~Pan, A.~Manjunath, T.~G. Rogers, T.~M. Aamodt, and N.~Hardavellas, ``{AccelWattch: A Power Modeling Framework for Modern {GPU}s},'' in \emph{MICRO-54: 54th Annual IEEE/ACM International Symposium on Microarchitecture}.\hskip 1em plus 0.5em minus 0.4em\relax New York, NY, USA: Association for Computing Machinery, 2021, p. 738–753. [Online]. Available: \url{https://doi.org/10.1145/3466752.3480063}
\BIBentrySTDinterwordspacing

\bibitem{powersensor3}
\BIBentryALTinterwordspacing
S.~van~der Vlugt, L.~Oostrum, G.~Schoonderbeek, B.~van Werkhoven, B.~Veenboer, K.~Doekemeijer, and J.~Romein, ``{PowerSensor3: A Fast and Accurate Open Source Power Measurement Tool},'' in \emph{Proceedings of the International Symposium on Performance Analysis of Systems and Software (ISPASS)}, 2025. [Online]. Available: \url{https://doi.org/10.48550/arXiv.2504.17883}
\BIBentrySTDinterwordspacing

\bibitem{ear}
J.~Corbal{\'a}n and L.~Brochard, ``{EAR: Energy management framework for HPC},'' \url{https://www.bsc.es/research-and-development/software-and-apps/software-list/ear-energy-management-framework-hpc}, 2018.

\bibitem{accelsim}
\BIBentryALTinterwordspacing
M.~Khairy, Z.~Shen, T.~M. Aamodt, and T.~G. Rogers, ``{Accel-Sim: An Extensible Simulation Framework for Validated {GPU} Modeling},'' in \emph{2020 ACM/IEEE 47th Annual International Symposium on Computer Architecture (ISCA)}, 2020, pp. 473--486. [Online]. Available: \url{https://doi.org/10.1109/ISCA45697.2020.00047}
\BIBentrySTDinterwordspacing

\bibitem{powerlog}
A.~R. Shovon, ``{Powerlog: Lightweight Power Profiling Tool for NVIDIA {GPU}s},'' \url{https://pypi.org/project/powerlog}, 2026.

\bibitem{likwid}
\BIBentryALTinterwordspacing
J.~Treibig, G.~Hager, and G.~Wellein, ``{LIKWID: A Lightweight Performance-Oriented Tool Suite for x86 Multicore Environments},'' in \emph{2010 39th International Conference on Parallel Processing Workshops}, 2010, pp. 207--216. [Online]. Available: \url{https://doi.org/10.1109/ICPPW.2010.38}
\BIBentrySTDinterwordspacing

\bibitem{hpctoolkit}
\BIBentryALTinterwordspacing
L.~Adhianto, S.~Banerjee, M.~Fagan, M.~Krentel, G.~Marin, J.~Mellor-Crummey, and N.~R. Tallent, ``{HPCTOOLKIT: tools for performance analysis of optimized parallel programs},'' \emph{Concurrency and Computation: Practice and Experience}, vol.~22, no.~6, pp. 685--701, 2010. [Online]. Available: \url{https://doi.org/10.1002/cpe.1553}
\BIBentrySTDinterwordspacing

\bibitem{watsonai}
\BIBentryALTinterwordspacing
H.~Huang, K.~Zhang, H.~Liao, K.~Wu, and G.~Tang, ``{AIMeter: Measuring, Analyzing, and Visualizing Energy and Carbon Footprint of {AI} Workloads},'' in \emph{ArXiv preprint}, 2025. [Online]. Available: \url{https://doi.org/10.48550/arXiv.2506.20535}
\BIBentrySTDinterwordspacing

\bibitem{wattscope}
\BIBentryALTinterwordspacing
X.~Guan, N.~Bashir, D.~Irwin, and P.~Shenoy, ``{WattScope: Non-intrusive Application-level Power Disaggregation in Datacenters},'' \emph{SIGMETRICS Perform. Eval. Rev.}, vol.~51, no.~4, p. 24–25, Feb. 2024. [Online]. Available: \url{https://doi.org/10.1145/3649477.3649491}
\BIBentrySTDinterwordspacing

\bibitem{benoit_courty_2024}
\BIBentryALTinterwordspacing
B.~Courty, V.~Schmidt, S.~Luccioni, Goyal-Kamal, MarionCoutarel, B.~Feld, J.~Lecourt, LiamConnell, A.~Saboni, Inimaz, supatomic, M.~Léval, L.~Blanche, A.~Cruveiller, ouminasara, F.~Zhao, A.~Joshi, A.~Bogroff, H.~de~Lavoreille, N.~Laskaris, E.~Abati, D.~Blank, Z.~Wang, A.~Catovic, M.~Alencon, M.~Stechly, C.~Bauer, L.~O.~N. de~Araújo, JPW, and MinervaBooks, ``{mlco2/codecarbon: v3.2.6},'' May 2026. [Online]. Available: \url{https://doi.org/10.5281/zenodo.19334697}
\BIBentrySTDinterwordspacing

\bibitem{koomey2007_tco}
J.~Koomey, K.~Brill, P.~Turner, J.~Stanley, and B.~Taylor, ``{A Simple Model for Determining True Total Cost of Ownership for Data Centers},'' \url{https://m.softchoice.com/files/pdf/about/sustain-enable/simplemodeldetermingtruetco.pdf}, 2007.

\bibitem{nvidia_tco}
NVIDIA, ``{Total Cost of Ownership ({TCO}) resources and calculators},'' \url{https://www.nvidia.com/en-us/networking/total-cost-ownership}, 2026.

\bibitem{intel_xeon_advisor}
Intel, ``{Intel {X}eon {P}rocessor {A}dvisor},'' \url{https://xeonprocessoradvisor.intel.com}, 2026.

\bibitem{amd_epyc_tco}
AMD, ``{AMD {EPYC} {S}erver {V}irtualization {TCO} Estimation Tool},'' \url{https://www.amd.com/en/processors/epyc-VirtTCOtool}, 2026.

\bibitem{scale_tco}
{Scale Computing}, ``{Total Cost of Ownership ({TCO}) Calculator},'' \url{https://www.scalecomputing.com/total-cost-of-ownership-tco-calculator}, 2026.

\bibitem{aws_calculator}
{Amazon Web Services (AWS)}, ``{AWS Pricing Calculator},'' \url{https://calculator.aws}, 2026.

\bibitem{google_calculator}
{Google Cloud}, ``{Google Cloud Pricing Calculator},'' \url{https://cloud.google.com/products/calculator}, 2026.

\bibitem{azure_calculator}
{Microsoft Azure}, ``{Azure Pricing Calculator},'' \url{https://azure.microsoft.com/en-us/pricing/calculator}, 2026.

\bibitem{cloudcarbonfootprint}
{ThoughtWorks and the Cloud Carbon Footprint community}, ``{Cloud Carbon Footprint},'' \url{https://www.cloudcarbonfootprint.org/} and \url{https://github.com/cloud-carbon-footprint/cloud-carbon-footprint}, 2026.

\bibitem{dcpro}
{Lawrence Berkeley National Laboratory (LBNL)}, ``{DC {P}ro: {D}ata {C}enter {P}rofiler},'' \url{https://datacenters.lbl.gov/dcpro}, 2026.

\bibitem{lttco}
\BIBentryALTinterwordspacing
W.~Yan, J.~Yao, Q.~Cao, and Y.~Zhang, ``{LT-TCO: A {TCO} Calculation Model of Data Centers for Long-Term Data Preservation},'' in \emph{2019 IEEE International Conference on Networking, Architecture and Storage (NAS)}, 2019, pp. 1--8. [Online]. Available: \url{https://doi.org/10.1109/NAS.2019.8834714}
\BIBentrySTDinterwordspacing

\bibitem{ipack2007}
\BIBentryALTinterwordspacing
C.~L. Belady and C.~G. Malone, ``{Metrics and an Infrastructure Model to Evaluate Data Center Efficiency},'' in \emph{{Proceedings of the ASME 2007 InterPACK Conference collocated with the ASME/JSME 2007 Thermal Engineering Heat Transfer Summer Conference}}, ser. International Electronic Packaging Technical Conference and Exhibition, vol.~1, 2007, pp. 751--755. [Online]. Available: \url{https://doi.org/10.1115/IPACK2007-33338}
\BIBentrySTDinterwordspacing

\bibitem{spie2007}
\BIBentryALTinterwordspacing
B.~Denisenko, M.~Tyanutov, I.~Nikiforov, and S.~Ustinov, ``{Algorithm for Calculating {TCO} and {SCE} Metrics to Assess the Efficiency of Using a Data Center},'' in \emph{2nd International Conference on Computer Applications for Management and Sustainable Development of Production and Industry (CMSD-II-2022)}, S.~Sadullozoda and A.~Gibadullin, Eds., vol. 12564, International Society for Optics and Photonics.\hskip 1em plus 0.5em minus 0.4em\relax SPIE, 2023, p. 1256403. [Online]. Available: \url{https://doi.org/10.1117/12.2669285}
\BIBentrySTDinterwordspacing

\bibitem{specpower}
{Standard Performance Evaluation Corporation ({SPEC})}, ``{SPEC Power Benchmark},'' \url{https://www.spec.org/power_ssj2008}, 2026.

\bibitem{specsert}
------, ``{SPEC SERT: Server Efficiency Rating Tool},'' \url{https://www.spec.org/sert}, 2026.

\bibitem{mlperfpower}
MLPerf, ``{MLPerf Power Benchmark},'' \url{https://mlperf.org/power}, 2026.

\bibitem{wadenstein2025lca}
\BIBentryALTinterwordspacing
M.~Wadenstein and W.~Vanderbauwhede, ``{Life Cycle Analysis for Emissions of Scientific Computing Centres},'' \emph{The European Physical Journal C}, vol.~85, p. 913, 2025. [Online]. Available: \url{https://doi.org/10.1140/epjc/s10052-025-14650-8}
\BIBentrySTDinterwordspacing

\bibitem{aritz2025}
\BIBentryALTinterwordspacing
P.~Arzt and F.~Wolf, ``{Navigating Energy Doldrums: Modeling the Impact of Energy Price Volatility on HPC Cost of Ownership},'' \emph{ArXiv preprint}, 2025. [Online]. Available: \url{https://arxiv.org/abs/2509.07567}
\BIBentrySTDinterwordspacing

\bibitem{AfzalThesis:2015}
\BIBentryALTinterwordspacing
A.~Afzal, ``{The cost of {computation}: {Metrics} and models for modern {multicore}-{based} systems in scientific computing},'' \emph{{Master}'s thesis, Department Informatik, Friedrich Alexander Universit{\"a}t Erlangen-N{\"u}rnberg}, 2015. [Online]. Available: \url{https://doi.org/10.13140/RG.2.2.35954.25283}
\BIBentrySTDinterwordspacing

\bibitem{Afzal:2023:2}
\BIBentryALTinterwordspacing
A.~Afzal, G.~Hager, and G.~Wellein, ``{{SPEChpc} {2021} Benchmarks on Ice Lake and Sapphire Rapids Infiniband Clusters{:} A Performance and Energy Case Study},'' in \emph{14th IEEE/ACM Performance Modeling, Benchmarking and Simulation of High Performance Computer Systems (PMBS)}, 2023. [Online]. Available: \url{https://doi.org/10.1145/3624062.3624197}
\BIBentrySTDinterwordspacing

\bibitem{AfzalHW:2025:1}
\BIBentryALTinterwordspacing
------, ``{Analytic Roofline Modeling and Energy Analysis of LULESH Proxy Application on Multi-Core Clusters},'' \emph{International Journal of High Performance Computing Applications (IJHPCA)}, 2025. [Online]. Available: \url{https://doi.org/10.1177/10943420251363711}
\BIBentrySTDinterwordspacing

\bibitem{nvidia_specs}
{NVIDIA Corporation}, ``{NVIDIA {GPU} Architecture Specifications and Datasheets},'' \url{https://www.nvidia.com/en-us/data-center}, 2026.

\bibitem{energystar}
{ENERGY STAR}, ``{ENERGY STAR Program for Data Center Equipment},'' \url{https://www.energystar.gov/products/data_center_equipment}, 2026.

\bibitem{AfzalHW:2025:2}
\BIBentryALTinterwordspacing
A.~Afzal, A.~Kahler, G.~Hager, and G.~Wellein, ``{GROMACS Unplugged: How Power Capping and Frequency Shapes Performance on GPUs},'' \emph{Euro-Par 2025: Parallel Processing Workshops Volume in the Springer Lecture Notes in Computer Science (LNCS) series}, 2025. [Online]. Available: \url{https://doi.org/10.48550/arXiv.2412.08792}
\BIBentrySTDinterwordspacing

\bibitem{dvfsmodel}
\BIBentryALTinterwordspacing
S.~Hong and H.~Kim, ``{An Integrated {GPU} Power and Performance Model},'' in \emph{Proceedings of the 37th Annual International Symposium on Computer Architecture (ISCA)}, 2010, pp. 280--289. [Online]. Available: \url{https://doi.org/10.1145/1815961.1815998}
\BIBentrySTDinterwordspacing

\bibitem{Amati:2025}
\BIBentryALTinterwordspacing
G.~Amati, M.~Turisini, A.~Monterubbiano, M.~Paladino, E.~Boella, D.~Gregori, and D.~Croce, ``{Experience on Clock Rate Adjustment for Energy-Efficient GPU-Accelerated Real-World Codes},'' in \emph{High Performance Computing}.\hskip 1em plus 0.5em minus 0.4em\relax Cham: Springer Nature Switzerland, 2026, pp. 245--257. [Online]. Available: \url{https://doi.org/10.1007/978-3-032-07612-0_19}
\BIBentrySTDinterwordspacing

\bibitem{Pall:2020}
\BIBentryALTinterwordspacing
S.~P{\'a}ll, A.~Zhmurov, P.~Bauer, M.~Abraham, M.~Lundborg, A.~Gray, B.~Hess, and E.~Lindahl, ``Heterogeneous parallelization and acceleration of molecular dynamics simulations in {GROMACS},'' \emph{The Journal of Chemical Physics}, vol. 153, no.~13, p. 134110, 2020. [Online]. Available: \url{https://doi.org/10.1063/5.0018516}
\BIBentrySTDinterwordspacing

\bibitem{SOBOL:2001}
\BIBentryALTinterwordspacing
I.~Sobol, ``{Global sensitivity indices for nonlinear mathematical models and their Monte Carlo estimates},'' \emph{Mathematics and Computers in Simulation}, vol.~55, no.~1, pp. 271--280, 2001. [Online]. Available: \url{https://doi.org/10.1016/S0378-4754(00)00270-6}
\BIBentrySTDinterwordspacing

\bibitem{Saltelli:2010}
\BIBentryALTinterwordspacing
A.~Saltelli, P.~Annoni, I.~Azzini, F.~Campolongo, M.~Ratto, and S.~Tarantola, ``{Variance based sensitivity analysis of model output. Design and estimator for the total sensitivity index},'' \emph{Computer Physics Communications}, vol. 181, no.~2, pp. 259--270, 2010. [Online]. Available: \url{https://doi.org/10.1016/j.cpc.2009.09.018}
\BIBentrySTDinterwordspacing

\bibitem{Afzal:2026sweet}
\BIBentryALTinterwordspacing
A.~Afzal, M.~M. Li, and M.~Panzlaff, ``Modeling and chasing the energy-efficiency sweet spots in modern {GPUs},'' in \emph{Parallel Processing and Applied Mathematics (PPAM 2026)}, ser. Lecture Notes in Computer Science.\hskip 1em plus 0.5em minus 0.4em\relax Springer, 2026. [Online]. Available: \url{https://doi.org/10.48550/arXiv.2607.00819}
\BIBentrySTDinterwordspacing

\bibitem{Ma:2026}
\BIBentryALTinterwordspacing
B.~Ma, A.~Afzal, J.~Eitzinger, and G.~Wellein, ``The illusion of power capping in {LLM} decode: A phase-aware energy characterisation across attention architectures,'' in \emph{Parallel Processing and Applied Mathematics (PPAM 2026)}, ser. Lecture Notes in Computer Science.\hskip 1em plus 0.5em minus 0.4em\relax Springer, 2026. [Online]. Available: \url{https://arxiv.org/abs/2605.11999}
\BIBentrySTDinterwordspacing

\bibitem{slurmenergyplugin}
``{Slurm Energy Plugin},'' \url{https://slurm.schedmd.com/slurm.conf.html}, 2026.

\end{thebibliography}


\begin{thebibliography}{99}
\section{Wattlytics Reproducibility Links}
\bibitem[S1]{web1} \url{https://tinyurl.com/Wattlytics-R1} 
\bibitem[S2]{web2} \url{https://tinyurl.com/Wattlytics-R2} 
\bibitem[S3]{web3} \url{https://tinyurl.com/Wattlytics-R3} 
\bibitem[S4]{web4} \url{https://tinyurl.com/Wattlytics-R4} 
\bibitem[S5]{web5} \url{https://tinyurl.com/Wattlytics-R5} 
\bibitem[S6]{web6} \url{https://tinyurl.com/Wattlytics-R6} 
\bibitem[S7]{web7} \url{https://tinyurl.com/Wattlytics-R7} 
\bibitem[S8]{web8} \url{https://tinyurl.com/Wattlytics-R8} 
\bibitem[S9]{web9} \url{https://tinyurl.com/Wattlytics-R9} 
\bibitem[S10]{web10} \url{https://tinyurl.com/Wattlytics-R10} 
\bibitem[S11]{web11} \url{https://tinyurl.com/Wattlytics-R11} 

\section{Table II Links}
\bibitem[I]{webI} 
\url{https://www.nvidia.com/es-la/data-center/l4}
\bibitem[II]{webII} 
\url{https://images.nvidia.com/content/Solutions/data-center/a40/nvidia-a40-datasheet.pdf}
\bibitem[III]{webIII} 
\url{https://images.nvidia.com/content/Solutions/data-center/vgpu-L40-datasheet.pdf}
\bibitem[IV]{webIV} 
\url{https://www.nvidia.com/de-de/data-center/a100}
\bibitem[V]{webV} 
\url{https://www.nvidia.com/de-de/data-center/h100}
\bibitem[VI]{webVI} 
\url{https://neobitti.com/wp-content/uploads/2024/07/NVIDIA-GH200-Grace-Hopper-Superchip.pdf}

\section{Other Links}
\bibitem[LI]{webLI} 
\url{https://ambermd.org/GPUPerformance.php}
\bibitem[LII]{webLII} 
\url{https://www.deltacomputer.com}
\bibitem[LIII]{webLIII} 
\url{https://github.com/pieroxy/lz-string}
\bibitem[LIV]{webLIV} 
\url{https://zenodo.org/records/23258567}

\end{thebibliography}

\renewcommand{\refname}{Sitography}
[Last accessed: Oct 8, 2026]

\appendices

\end{document}